\documentclass[superscriptaddress,
twocolumn,showpacs,preprintnumbers,amsmath,amssymb]{revtex4}
\usepackage{graphicx}
\usepackage{dcolumn}
\usepackage{bm}
\usepackage{latexsym}
\begin{document}
\title{Stochastic kinetics of ribosomes: single motor properties and collective behavior} 
\author{Ashok Garai}
\author{Debanjan Chowdhury}
\author{Debashish Chowdhury{\footnote{Corresponding author(E-mail: debch@iitk.ac.in)}}}
\affiliation{Department of Physics, Indian Institute of Technology,
Kanpur 208016, India.}
\author{T.V. Ramakrishnan}
\affiliation{Department of Physics, Banaras Hindu University, Varanasi 221005, India.}
\affiliation{Department of Physics, Indian Institute of Science, Bangalore 560012, India.}
\begin{abstract}
Synthesis of protein molecules in a cell are carried out by ribosomes.
A ribosome can be regarded as a molecular motor which utilizes the
input chemical energy to move on a messenger RNA (mRNA) track that
also serves as a template for the polymerization of the corresponding
protein. The forward movement, however, is characterized by an
alternating sequence of translocation and pause. Using a quantitative
model, which captures the mechanochemical cycle of an individual
ribosome, we derive an {\it exact} analytical expression for the
distribution of its dwell times at the successive positions on the
mRNA track. Inverse of the average dwell time satisfies a
``Michaelis-Menten-like'' equation and is consistent with the general
formula for the average velocity of a molecular motor with an unbranched
mechano-chemical cycle. Extending this formula appropriately, we also
derive the exact force-velocity relation for a ribosome. Often many
ribosomes simultaneously move on the same mRNA track, while each
synthesizes a copy of the same protein. We extend the model of a single
ribosome by incorporating steric exclusion of different individuals
on the same track. We draw the phase diagram of this model of ribosome
traffic in 3-dimensional spaces spanned by experimentally controllable
parameters. We suggest new experimental tests of our theoretical predictions.
\end{abstract}
\pacs{87.16.Ac  89.20.-a}
\maketitle
\section{Introduction}

Ribosome is one of the largest and most complex intracellular cyclic 
molecular machines \cite{spirinbook,spirin02,abel96,frank06}  
and it plays a crucial role in gene expression \cite{cellbook}. 
It synthesizes a protein molecule, which is a hetero polymer of amino 
acid subunits, using a messenger RNA (mRNA) as the corresponding 
template; this process is called {\it translation} (of the genetic 
message). Monomeric subunits of RNA are nucleotides and triplets of 
nucleotides constitute a codon. The dictionary of translation relates 
each type of possible codon with one species of amino acid. Thus, the 
sequence of amino acids on a protein is dictated by the sequence of 
codons on the corresponding template mRNA. The polymerization of protein 
takes places in three stages which are identified as {\it initiation}, 
{\it elongation} (of the protein) and {\it termination}. In this paper 
we focus almost exclusively on the elongation stage.

A ribosome is often treated as a molecular motor for which the mRNA 
template also serves as a track. In each step it moves forward on its 
track by one codon by consuming chemical fuel [e.g., two guanosine 
tri-phosphate(GTP) molecules]. Simultaneously, in each step, it also 
elongates the protein by adding an amino acid; the correct sequence 
of the amino acids required for polymerizing a protein is dictated by 
the codon sequence on the mRNA template. Therefore, it may be more 
appropriate to regard a ribosome as a mobile workshop that provides 
a platform for operation of several tools in a well coordinated manner. 
Our main aim is to predict the effects of the mechano-chemical cycle 
of individual ribosomes, in the elongation stage, on their experimentally 
measurable physical properties.  We first focus on the single-ribosome 
properties which characterize their stochastic movement on the track 
in the absence of inter-ribosome interactions. Then we consider the 
additional effects of the steric interactions of the ribosomes and 
those of the rates of initiation and termination of translation on the 
collective spatio-temporal organization of the ribosomes on a track.

The stochastic forward movement of a ribosome is characterized by an 
alternating sequence of pause and translocation. The sum of the 
durations of a pause and the following translocation defines the time 
of a dwell at the corresponding codon. Recently, using an ingenious 
method, the distribution $f(t)$ of the dwell times of a ribosome has 
been measured \cite{wen08}. We present a systematic derivation of this 
distribution from a detailed kinetic theory of translation which 
incorporates the mechano-chemical cycle of individual ribosomes.

The {\it exact} analytical expression for $f(t)$ which we derive here  
is, in general, a superposition of several exponentials. On the other 
hand, it has been claimed \cite{wen08} that difference of two 
exponentials fit the experimentally measured $f(t)$ very well. We  
reconcile these two observations by identifying the parameter regime 
where our theoretically derived $f(t)$ is, indeed, well approximated 
by difference of two exponentials 
\cite{redner,chemla08,shaevitz05,liao07,linden07}. Moreover, we show that 
$\langle t \rangle^{-1}$, inverse of the mean dwell time, satisfies a 
{\it Michaelis-Menten-like equation} \cite{dixon}. The reason for this 
feature of the mean dwell time is traced to the close formal similarity 
between the mechanochemical cycle of a ribosome and the catalytic cycle 
in the Michaelis-Menten theory of enzymes \cite{dixon}. 

The elongation of the growing protein by one amino acid is coupled 
to the translocation of the ribosome by one codon. Therefore,  
$\langle t \rangle^{-1}$ is also the average velocity 
$\langle V \rangle$ of a ribosome on the mRNA track. An analytical 
expression for the average velocity of a molecular motor, whose 
mechano-chemical cycle is unbranched, was derived by Fisher and 
Kolomeisky \cite{fishkolo} in the context of motors involved in 
intracellular transport of cargoes \cite{kolorev}. The mechano-chemical 
cycle of the ribosome in our model is, at least formally, a special 
case of the cycle in the Fisher-Kolomeisky model. In this special case, 
the Fisher-Kolomeisky formula for the average velocity of the molecular 
motor, indeed, reduces to the expression for $\langle t \rangle^{-1}$ 
in our model of ribosome. 

The average velocity of a ribosome can be reduced also by applying 
an external force (called a load force) that opposes the natural 
movement of the ribosome on its track. The force-velocity relation 
$\langle V \rangle(F)$ (i.e., the variation of the average velocity 
$\langle V \rangle$ of a motor with increasing load force $F$) is 
one of the most important characteristics of a molecular motor. 
Inspired by the recent progress in the single-ribosome imaging and 
manipulation techniques 
\cite{marshall08,blanchard09,blanchard04,uemura07,munro08,vanzi07,wang07c,wen08}, 
we extend the formula for $\langle t \rangle^{-1}$ appropriately to 
derive $\langle V \rangle(F)$ for single ribosomes. The smallest 
load force which is just adequate to stall a molecular motor on its 
track is called the stall force $F_{s}$. We also predict the 
dependence of $F_{s}$ on the availability of the amino acid monomers 
and the concentration of GTP molecules. 

Our theoretical predictions for $f(t)$, $\langle V \rangle(F)$ and 
$F_{s}$ show explicitly how these quantities depend on various 
experimentally controllable parameters. Deep understanding of these 
dependences will also help in controlling various features of $f(t)$, 
$\langle V \rangle(F)$ and $F_{s}$. In principle, the validity and 
accuracy of our theoretical predictions can be tested by repeating 
{\it in-vitro} experiments of ref.\cite{wen08} for several different 
concentrations of the amino acid monomers and GTP molecules. 

Often many ribosomes simultaneously move on the same mRNA track, while 
each synthesizes separately a copy of the same protein. We refer to 
such collective movement of ribosomes on a mRNA strand as ribosome
traffic because of its superficial similarity with vehicular traffic
\cite{polrev}. In most of the earlier theoretical studies of ribosome
traffic, individual ribosomes have been modelled as hard rods and
their steric interactions have been captured by mutual exclusion
\cite{macdonald68,macdonald69,lakatos03,shaw03,shaw04a,shaw04b,chou03,chou04,dong1,dong2}. 
Thus, all those models may be regarded as totally asymmetric simple
exclusion process (TASEP) for hard rods \cite{sz,schuetz}.
In some recent works \cite{bcajp,bcpre} we have extended these
TASEP-type models of ribosome traffic by capturing the essential steps
of the mechano-chemical cycle of individual ribosomes. We have also 
reported the variation of the average rate of protein synthesis with 
increasing population density of the ribosomes on the track. In this 
work we present the phase diagrams of the model of ribosome traffic. 

In the earlier TASEP-type models of ribosome traffic
\cite{macdonald68,macdonald69,lakatos03,shaw03,shaw04a,shaw04b,chou03,chou04,dong1,dong2}
, the phase diagrams were plotted in a two-dimensional plane spanned by
$\alpha$ and $\beta$, which determine the rates of initiation and termination.
In this paper we plot the  three-dimensional phase diagrams of our model
of ribosome traffic  \cite{bcpre} in spaces spanned by three parameters
which, for different diagrams, are selected from $\alpha$, $\beta$, the
availability of amino acid monomers and the rate of GTP hydrolysis.
Compared to the two-dimensional phase diagram of the TASEP-type models
of ribosome traffic, these three-dimensional phase diagrams provide
deeper insight into the interplay of single ribosome-mechanochemistry
and their collective spatio-temporal organization.

Traffic-like collective movements of ribosomes on a mRNA track during 
translation of a gene was demonstrated many years ago by electron 
microscopy \cite{phystoday}. However, to our knowledge, no attempt has 
been made so far to study the phase diagram of ribosome traffic by 
systematic quantitative measurements. But, in contrast to most of the 
earlier works, we have used experimentally controllable parameters to 
plot the phase diagrams. Therefore, we hope, this paper will stimulate 
experimental studies of the phase diagrams by systematically varying 
the supply of amino acids (monomeric subunits of protein) and GTP 
molecules (fuel of ribosomes) in the solution.

The paper is organized as follows: In section \ref{model} we introduce 
the model of the mechano-chemical cycle of individual ribosomes. The 
exact dwell time distribution is calculated in section \ref{dwell}, 
while the mean dwell time and the physical interpretations of the 
Michaelis-Menten-like equation are presented in section \ref{meandwell}. 
The connection between the mean dwell time and average velocity of a 
ribosome are pointed out in section \ref{fvrel}, where we also show 
the trends of variation of the force-velocity relation with variation 
of some key parameters of the model. The variance of the dwell time 
distribution and the diffusion constant of a ribosome are quantitative 
measures of fluctuations; the analytical expressions of these quantities 
are presented in section \ref{fluc} where their relationships are pointed 
out. The distribution of the run times of the ribosomes on their 
track and the relation of its first two moments with the corresponding 
moments of the dwell time distribution are discussed in section 
\ref{run}. The effects of steric interactions among the ribosomes 
during their traffic-like collective movement on a single mRNA track 
are studied in section \ref{interaction}; the overall rate of protein 
synthesis are presented in subsection \ref{flux} while in subsection 
\ref{phase} we plot the 3d phase diagrams of the model and also depict 
2d projections to compare with the corresponding 2d phase diagrams of 
the TASEP. Finally, the results are summarized and main conclusions are 
drawn in section \ref{summary}.

\section{Model of mechano-chemical cycle of ribosome}\label{model}

\begin{figure}
\begin{center} 
(a)\\
\includegraphics[angle=-90,width=0.9\columnwidth]{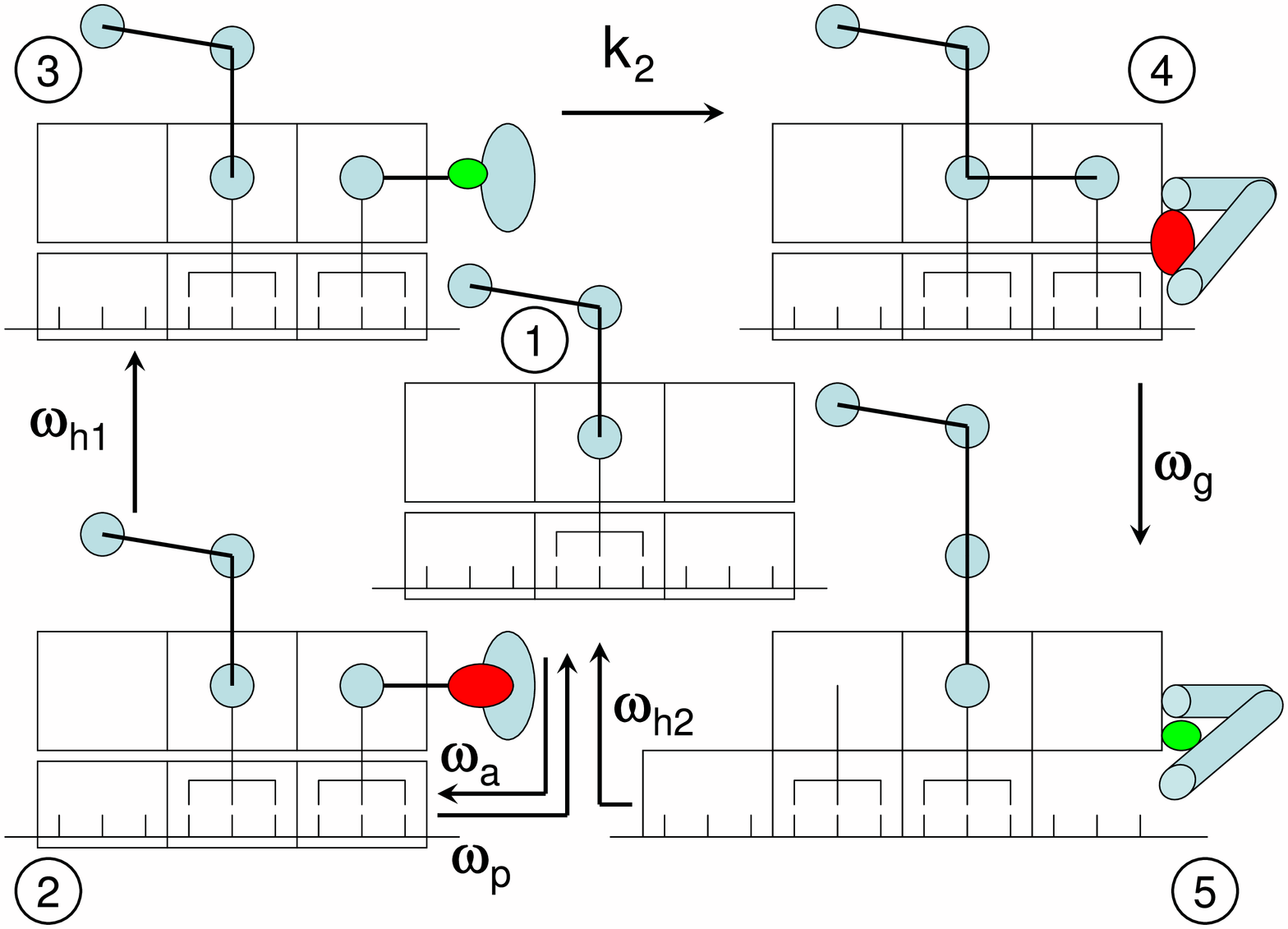} \\
\vspace{1cm} 
(b)\\
\includegraphics[angle=-90,width=0.9\columnwidth]{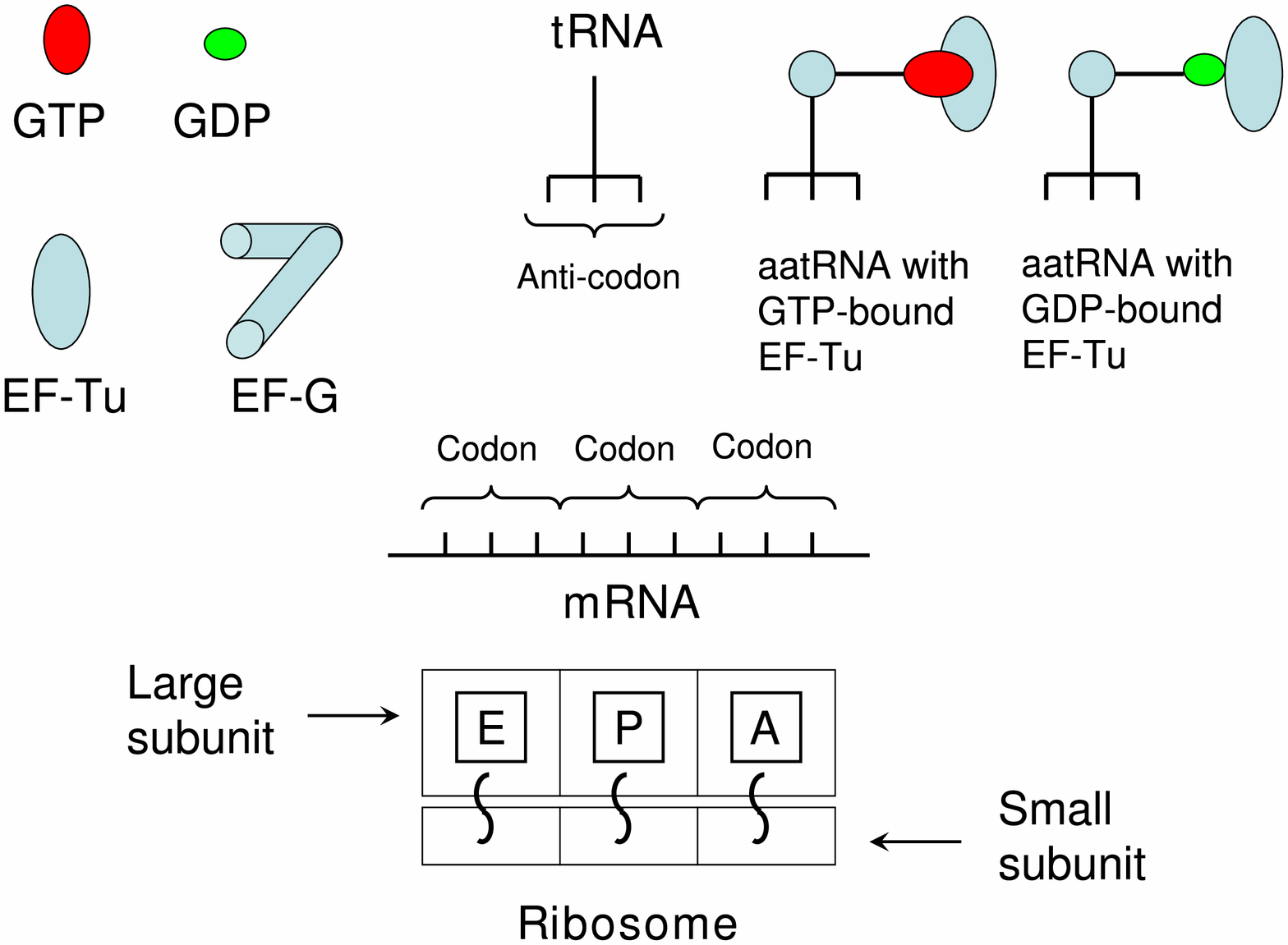}
\end{center}
\caption{(Color online) (a) A cartoon for pictorial depiction of the 
mechano-chemical cycle of an individual ribosome in our model. 
Some of the symbols are explained in (b).
}
\label{fig-fullcycle}
\end{figure}

\begin{figure}
\begin{center}
\includegraphics[angle=-90,width=0.9\columnwidth]{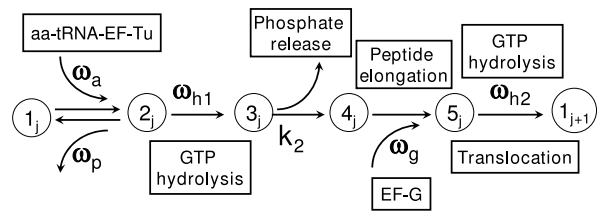}
\end{center}
\caption{Mechano-chemical cycle of an individual ribosome shown in 
fig.\ref{fig-fullcycle} is redrawn for the convenience of formulation 
of the master equations. 
}
\label{fig-model}
\end{figure}

The fig.\ref{fig-fullcycle} depicts the mechano-chemical cycle of each
ribosome in the stage of elongation of the protein where the integer 
index $j$ labels the codons on the mRNA track.  The amino acid 
monomers are supplied to the ribosome in a form in which they 
form a complex with an adapter molecule called tRNA; 
the complex is called aminoacyl-tRNA (aatRNA). Each ``charged'' 
aatRNA, bound to another protein called EF-Tu, arrives at the 
ribosome from the surrounding medium. The arrival of the correct 
aatRNA-EF-Tu, as dictated by the mRNA template, and its recognition 
by the ribosome located at the site $j$ triggers transition from 
the chemical state 1 to 2 in the same location with a transition 
rate $\omega_a$. However, if the aatRNA does not belong to the 
correct species, it is rejected, thereby causing the reverse transition 
from state 2 to state 1 with transition rate $\omega_p$. Hydrolysis 
of GTP drives the transition from state 2 to state 3 with the 
corresponding rate $\omega_{h1}$. Release of the phosphate group, 
a product of the GTP hydrolysis, is responsible for the transition 
from state 3 to state 4; the corresponding rate constant is $k_2$. 
The peptide bond formation between the newly arrived amino acid 
monomer and the growing protein, which leads to the elongation of 
the protein by one amino acid monomer, (and some associated biochemical 
processes, including the arrival of the protein EF-G), is captured by 
the next transition to the state 5 with transition rate $\omega_g$. 
All the subsequent processes, including the forward translocation 
of the ribosome by one codon, driven by the hydrolysis of another 
GTP molecule, and the exit of a naked tRNA from the ribosome complex 
are captured by a single effective transition from state 5 at site 
$j$ to the state $1$ at the site $j+1$ with the transition rate 
$\omega_{h2}$. The esssential processes of the cycle are summarized 
in the simplified figure \ref{fig-model}. More detailed explanations 
of the states and the transitions are given in ref.\cite{bcpre}.

\section{Dwell time distribution for a single ribosome: most general case}\label{dwell}

Because of recent improvements in experimental techniques, it has 
become possible to image and manipulate single ribosomes 
\cite{marshall08,blanchard09,blanchard04,uemura07,munro08,vanzi07,wang07c,wen08}.
In the recent experiments on single-ribosome manipulation \cite{wen08}, 
the distribution of the dwell times of a single ribosome at a codon 
was measured. It was also shown that the experimental data   
fit best to a difference of two exponentials. More recently, we 
\cite{gcr09} have demonstrated that the numerical data obtained from 
computer simulations of our model can also be fitted to a difference 
of two exponentials. In this section we derive an exact analytical 
formula for the dwell time distribution in our model and compare it 
with the corresponding numerical data obtained from computer simulations. 
This analytical formula shows how the distribution of the dwell times 
can be controlled by tuning the rates of the various sub-steps of a 
mechanochemical cycle of the ribosome. This is a new prediction which, 
in principle, can be tested by repeating the {\it in-vitro} single 
ribosome experiments \cite{wen08} for different concentrations of GTP 
and aa-tRNA molecules.

For every ribosome, the dwell time is measured by an imaginary 
``stopwatch'' which is reset to zero whenever the ribosome reaches 
the chemical state $1$, {\it for the first time}, after arriving at 
a new codon (say, $j+1$-th codon from the $j$-th codon). For the 
convenience of mathematical formulation, and for later comparison 
with the corresponding results of single molecule enzymatic kinetics, 
we make the following assumption: a ribosome finds itself in an 
excited state $1^{\ast}$ following the transition from the state 
$(j,5)$ to $(j+1,1^{\ast})$ and, then, relaxes to its normal state 
$(j+1,1)$ with a rate constant $\delta$. If the ribosome relaxes 
very rapidly from the state $1^{\ast}$ to the state $1$, we can 
set $\delta \to \infty$ at the end of the calculation. 

Let $P_{\mu}(j,t)$ be the probability of finding a ribosome at site 
$j$, in the chemical state $\mu$ at time $t$. For our calculations 
in this section, we do not need to write the site index $j$ explicitly. 
The time evolution of the probabilities $P_{\mu}(t)$ are given by 
\begin{equation}
 \frac{dP_1(t)}{dt}=-\omega_a P_1(t)+\omega_p P_2(t)
\label{eq-m1}
\end{equation}
\begin{equation}
 \frac{dP_2(t)}{dt}=\omega_a P_1(t)-(\omega_p+\omega_{h1}) P_2(t)
\label{eq-m2}
\end{equation}
\begin{equation}
 \frac{dP_3(t)}{dt}=\omega_{h1} P_2(t)-k_2 P_3(t)
\label{eq-m3}
\end{equation}
\begin{equation}
 \frac{dP_4(t)}{dt}=k_2 P_3(t)-\omega_g P_4(t)
\label{eq-m4}
\end{equation}
\begin{equation}
 \frac{dP_5(t)}{dt}=\omega_{g} P_4(t)-\omega_{h2} P_5(t)
\label{eq-m5}
\end{equation}
\begin{equation}
 \frac{dP_{1^{\ast}}(t)}{dt}=\omega_{h2} P_5(t)
\label{eq-m6}
\end{equation}

The probability that addition of a new amino acid subunit to the growing 
protein is completed between times $t$ and $t+\Delta t$ is $f(t) \Delta t$. 
But, 
\begin{equation}
f(t) \Delta t = \Delta P_{1^{\ast}}(t) = \omega_{h2} P_{5}(t) \Delta t
\label{eq-defft1}
\end{equation}
where $\Delta P_{1^{\ast}}(t)$ is the probability that the ribosome is in the 
state $1^{\ast}$ in the time interval between $t$ and $t+\Delta t$. Therefore, 
\begin{equation}
f(t) = \frac{dP_{1^{\ast}}(t)}{dt} = \omega_{h2} P_{5}(t) 
\label{eq-defft2}
\end{equation}

\begin{figure}
\begin{center} 
(a)\\
 \includegraphics[width=0.8\columnwidth]{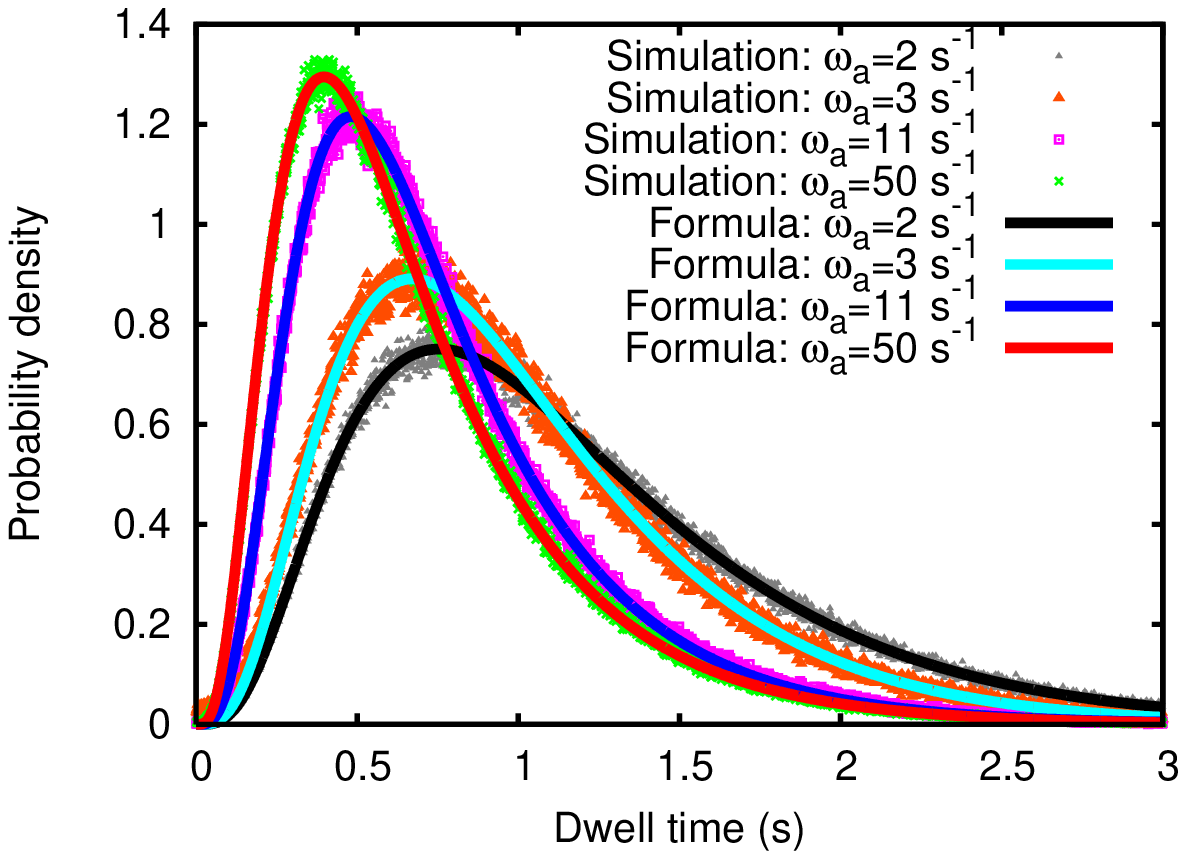} \\
\vspace{1cm}
(b)\\
 \includegraphics[width=0.8\columnwidth]{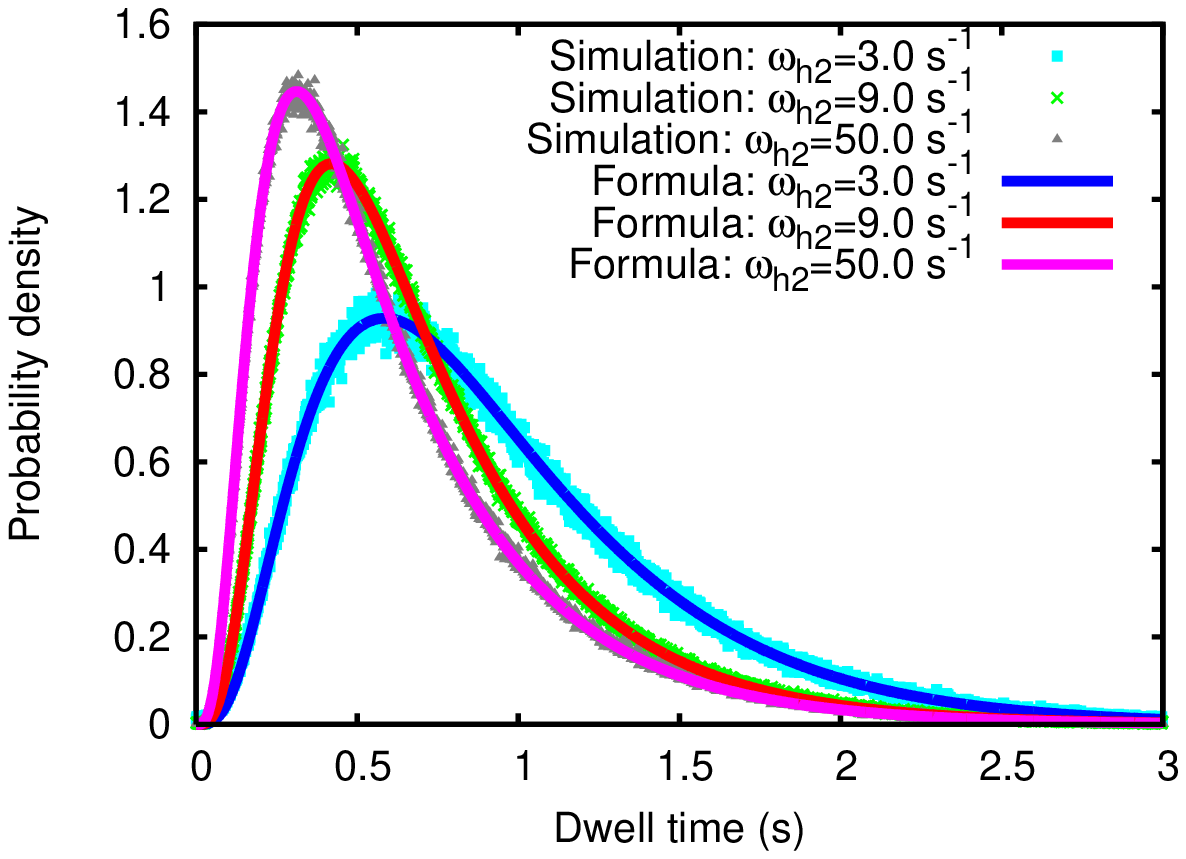}
\end{center}
\caption{(Color online) Probability density $f(t)$ of the dwell times 
of a single ribosome in the most general case of our model for a few 
different valus of (a) the parameter $\omega_{a}$ (which is proportional 
to the concentration of tRNA-bound amino acid subunits), and (b) 
the parameter $\omega_{h2}$ (which determines the rate of ``stepping''). 
The continuous curve corresponds to the analytically derived expression 
(\ref{eq-ftgen}) whereas the discrete data points have been obtained 
from computer simulation of the same model.}
\label{fig-ftgen}
\end{figure}

Solving the equations (\ref{eq-m1})-(\ref{eq-m6}), subject to the 
normalization condition 
\begin{equation}
 P_1(t)+P_2(t)+P_3(t)+P_4(t)+P_5(t)+P_{1^{\ast}}(t)=1,
\end{equation}
and the initial conditions 
\begin{equation}
P_1(0)=1, P_2(0)=P_3(0)=P_4(0)=P_5(0)=P_{1^{\ast}}(0)=0,
\label{6}
\end{equation}
we get the time-dependent probabilities $P_{\mu}(t)$ ($\mu=1,2,...,5$); 
the details are given in the appendix. Then, using the relation 
(\ref{eq-defft2}), we obtain the distribution $f(t)$ of the dwell times 
to be 
\begin{eqnarray}
&&f(t) = C_{1}exp(-\omega_{1}t) + C_{2}exp(-\omega_{2}t) \nonumber \\ 
&+& C_{3} exp(-k_2t)+C_{4}exp(-\omega_gt)+C_{5}exp(-\omega_{h2}t) \nonumber \\
\label{eq-ftgen}
\end{eqnarray}
where 
\begin{equation}
 C_1=\frac{\omega_a\omega_{h1}k_2\omega_g\omega_{h2}}{(\omega_2-\omega_1)(k_2-\omega_1)(\omega_g-\omega_1)(\omega_{h2}-\omega_1)}
\end{equation}
\begin{equation}
 C_2=\frac{\omega_a\omega_{h1}k_2\omega_g\omega_{h2}}{(\omega_1-\omega_2)(k_2-\omega_2)(\omega_g-\omega_2)(\omega_{h2}-\omega_2)}
\end{equation}
\begin{equation}
 C_3=\frac{\omega_a\omega_{h1}k_2\omega_g\omega_{h2}}{(\omega_1-k_2)(\omega_2-k_2)(\omega_g-k_2)(\omega_{h2}-k_2)}
\end{equation}
\begin{equation}
 C_4=\frac{\omega_a\omega_{h1}k_2\omega_g\omega_{h2}}{(\omega_1-\omega_g)(\omega_2-\omega_g)(k_2-\omega_g)(\omega_{h2}-\omega_g)} 
\end{equation}
\begin{equation}
 C_5=\frac{\omega_a\omega_{h1}k_2\omega_g\omega_{h2}}{(\omega_1-\omega_{h2})(\omega_2-\omega_{h2})(k_2-\omega_{h2})(\omega_{g}-\omega_{h2})} 
\end{equation}
and
\begin{equation}
\omega_{1} = \frac{\omega_{h1}+\omega_p+\omega_a}{2} - [\sqrt{\frac{(\omega_{h1}+\omega_p+\omega_a)^2}{4}-\omega_a\omega_{h1}}]
\label{eq-omega1}
\end{equation}
\begin{equation}
\omega_{2} = \frac{\omega_{h1}+\omega_p+\omega_a}{2} + [\sqrt{\frac{(\omega_{h1}+\omega_p+\omega_a)^2}{4}-\omega_a\omega_{h1}}] 
\label{eq-omega2}
\end{equation}
The explicit mathematical formula (\ref{eq-ftgen}) for the dwell time
distribution, which we report in this manuscript, predicts how the
distribution depends quantitatively on the rates of the steps in the
mechanochemical cycle of a ribosome. These predictions can be tested
by repeating the experiments of Wen et al. \cite{wen08} with different 
concentrations of amino-acid subunits of the proteins (i.e., aatRNA 
molecules), fuel of ribosome motor (i.e., GTP molecules) and ribosomes.

We plot the distribution (\ref{eq-ftgen}) in fig.\ref{fig-ftgen} and 
compare it with the corresponding distribution which we have obtained 
by direct computer simulation of our model. The agreement between the 
theoretical formula (\ref{eq-ftgen}) and the simulation data is excellent. 
Note that $\sum_{\mu=1}^{5} C_{\mu} = 0$, which implies that $f(t) = 0$ 
at $t=0$. Moreover, the nonmonotonic variation of $f(t)$ with $t$ 
arises from the fact that not all of the coefficients $C_{\mu}$ are 
positive. As $\omega_a$ {\it decreases} (i.e., effectively, aatRNA 
become more scarce), the tail of the distribution becomes longer and 
the peak shifts to longer dwell times. Moreover, similar trend is 
observed also in the variation of the most probable dwell time with 
the decrease of $\omega_{h2}$. The trend of variation of the width 
of the distribution will be discussed later in section{\ref{fluc}} 
of this paper.

\subsection{Special case I: $\omega_p=0$}

In the special case $\omega_p = 0$, 
\begin{eqnarray}
&& f(t) = C^{'}_1 exp(-\omega_a t) + C^{'}_2 exp(-\omega_{h1}t) \nonumber \\ 
&+& C^{'}_3 exp(-k_2t) + C^{'}_4 exp(-\omega_g t) + C^{'}_5 exp(-\omega_{h2}t) \nonumber \\
\label{eq-ftwp0}
\end{eqnarray}
where, $C^{'}_{\mu}$ is obtained from $C_{\mu}$ by replacing 
$\omega_{1}$ and $\omega_{2}$ by $\omega_{a}$ and $\omega_{h1}$, 
respectively. The form of the expression (\ref{eq-ftwp0}) of $f(t)$ 
makes the underlying physics very transparent- $f(t)$ is a 
superposition of five different terms each of which decays 
exponentially with one of the five rate constants. Moreover, a 
clear pattern in the factors in the denominators of the coefficients 
$C^{'}_{\mu}$  ($\mu = 1,2,...,5$) has also emerged.

\subsection{Special case II: $\omega_a=\omega_{h1}=k_2=\omega_g=\omega_{h2}, \omega_{p}=0$}

Note that we have derived the general expression (\ref{eq-ftgen}) for 
$f(t)$ assuming that no two rate constants are equal. One can envisage 
several different possible situations where two or more rate constants 
have identical numerical values \cite{garaithesis}. In order to 
demonstrate that the form of $f(t)$ can get modified under such special 
conditions, in this subsection we consider a very special case where 
$\omega_{p} = 0$ and all the nonvanishing rate constants are equal, i.e., 
$\omega_a=\omega_{h1}=\omega_{h2}=\omega_g=k_2=g$. 
In this case the master equations become much simpler and the expression 
for $f(t)$ simplifies to the Gamma distribution 
\begin{equation}
f(t)=\frac{g^k ~t^{k-1}e^{-g~ t}}{\Gamma(k)}
\label{eq-ftgf}
\end{equation}
where $\Gamma(k)$ is the Gamma function with $k=5$. 

\section{Mean dwell time: Michaelis-Menten equation?}\label{meandwell} 

\begin{figure}
 \begin{center}
 \includegraphics[angle=-90,width=0.8\columnwidth]{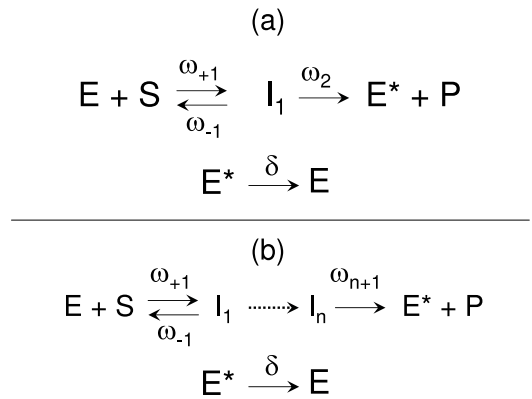}
\end{center}
\caption{(a) Catalytic cycle of an enzyme in the Michaelis-Menten theory. 
$E$ denotes the enzyme while $S$ and $P$ denote the substrate and product, 
respectively. The symbol $I_{1}$ represents the intermediate state of 
molecular complex of which the enzyme is a component. 
(b) Generalization of the cycle shown in (a) to $n$ number of intermediate 
states $I_{1},...,I_{n}$.}
\label{fig-MMreac}
\end{figure}

\begin{figure}
 \begin{center}
 \includegraphics[angle=-90,width=0.8\columnwidth]{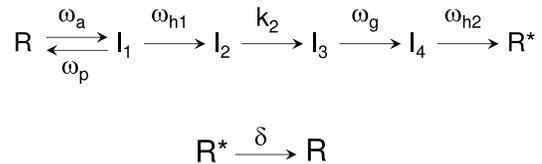}
\end{center}
\caption{The mechano-chemical cycle of a ribosome, shown in the 
fig.\ref{fig-model}, is redrawn for the convenience of comparison 
with the MM enzymatic reaction scheme shown in fig.\ref{fig-MMreac}(b).
The symbols $I_1, I_2, I_3, I_4$ denote the five intermediate states 
which we labelled in fig.\ref{fig-model} by the integers $1,2,3,4$, 
respectively.
}
\label{fig-RiboMM}
\end{figure}

\begin{figure}
 \begin{center}
 \includegraphics[angle=-90,width=0.8\columnwidth]{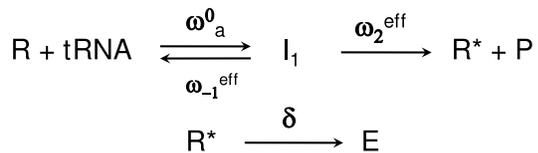}
\end{center}
\caption{The effective mechano-chemical cycle of a ribosome, where 
the effective rate constants $\omega_{2}^{eff}$ and $\omega_{-1}^{eff}$  
are given by the equations (\ref{eq-vmax}) and (\ref{eq-wmn1eff}), 
respectively. 
}
\label{fig-riboeffMM}
\end{figure}

Using the expresion for $f(t)$ in 
\begin{equation}
 \langle t \rangle=\int^\infty_0 tf(t)dt
\end{equation}
We get the mean dwell time 
\begin{equation}
\langle t \rangle=\frac{C_1}{\omega_1^2}+\frac{C_2}{\omega_2^2}+\frac{C_3}{k_2^2}+\frac{C_4}{\omega_g^2}+\frac{C_5}{\omega_{h2}^2}
\end{equation}
Further simplification gives,
\begin{equation}
 \langle t \rangle= \frac{1}{\omega_a}\biggl(1 + \frac{\omega_p}{\omega_{h1}}\biggr) +\frac{1}{\omega_{h1}} +\frac{1}{k_2} +\frac{1}{\omega_g} +\frac{1}{\omega_{h2}}
\label{mm-1}
\end{equation} 
which is, indeed, the sum of the average time periods spent in different 
steps of the mechano-chemical cycle.

Next we express the ``pseudo'' first order rate constant $\omega_{a}$ 
as $\omega_{a} = \omega_{a}^{0} [tRNA]$, 
where $[tRNA]$ is the concentration of the tRNA molecules. Then,  
the eqn. (\ref{mm-1}) can be recast as  
\begin{equation}
\langle t \rangle = \frac{1}{V_{max}}+\left( \frac{K_M}{V_{max}}\right)\frac{1}{[tRNA]}
\label{eq-effMM}
\end{equation}
where 
\begin{equation}
 \frac{1}{V_{max}}=\frac{1}{\omega_{h1}}+\frac{1}{k_2}+\frac{1}{\omega_g}+\frac{1}{\omega_{h2}} = \frac{1}{\omega_{2}^{eff}} 
\label{eq-vmax}
\end{equation} 
and 
\begin{equation}
K_M=\frac{\omega_{2}^{eff}+\omega_{-1}^{eff}}{\omega_a^{0}}
\label{eq-km}
\end{equation} 
with 
\begin{equation}
\omega_{-1}^{eff} = \omega_{p}\biggl(\frac{\omega_{2}^{eff}}{\omega_{h1}}\biggr)
\label{eq-wmn1eff}
\end{equation}

One remarkable feature of the expression (\ref{eq-effMM}) is that 
it is very similar to the Michaelis-Menten equation (MM equation) 
for the speed of enzymatic reactions in bulk \cite{dixon}. In 
chemical kinetics the MM equation is derived for the enzymatic 
cycle shown in fig.\ref{fig-MMreac} where the enzyme $E$ enhances 
the rate of the reaction that converts the substrates $S$ into the 
product $P$. In that case the maximum rate of the reaction is given 
by $V_{max} = \omega_{2}$ while the Michaelis constant is 
$(\omega_{+2}+\omega_{-1})/\omega_{+1}$.

The steps of the mechano-chemical cycle of an individual ribosome, 
as re-drawn in fig.\ref{fig-RiboMM}, are very similar to those of 
the generalized MM-like enzymatic cycle shown in fig.\ref{fig-MMreac}(b). 
The fact that the mean dwell time for the ribosomes follows a MM-like 
equation is consistent with the experimental observations in recent  
years \cite{qian02,english06,kou05,min05,min06,basu09}
that the average rate of an enzymatic reaction catalyzed by a single 
enzyme molecule is, most often, given by the same MM equation.

For our model, we can interpret $1/\langle t \rangle$ as the average 
rate at which a protein is synthesized by a ribosome, where aatRNA plays 
the role of the substrate and the protein elongated by one amino acid 
is the product. In the limit of effectively infinite supply of tRNA 
molecules, on the average, time required to complete one cycle would 
be the sum of the times required to complete the remaining steps of 
the cycle each of which has been assumed to be completely irreversible. 
This intuitive expectation for the maximum speed of protein synthesis 
is consistent with the analytical form (\ref{eq-vmax}) of $V_{max}$.  
Furthermore, in the expression (\ref{eq-km}) for the Michaelis constant 
the effective rate constants $\omega_{-1}^{eff}$ and $\omega_{2}^{eff}$ 
are the counterparts of $\omega_{-1}$ and $\omega_2$, respectively, 
of fig.\ref{fig-MMreac}(a). Therefore, so far as the average speed is 
concerned, the actual mechano-cycle, shown in fig.\ref{fig-RiboMM}, for 
a single ribosome can be replaced by the simpler MM-like cycle shown 
in fig.\ref{fig-riboeffMM} where $\omega_{a}^{0}$ is the counterpart 
of $\omega_{+1}$, In the limit 
$k_{2} \to \infty, \omega_{g} \to \infty, \omega_{h2} \to \infty$, 
the mechano-chemical cycle of a ribosome in our model reduces to the 
enzymatic cycle shown in fig.\ref{fig-MMreac}(a). In this limit, 
$\omega_{2}^{eff} \to \omega_{h1}$ and, hence, the expressions 
(\ref{eq-vmax}) and (\ref{eq-km}) reduce to the corresponding 
expressions for $V_{max}$ and $K_M$ in the MM equation for enzymes.

In reality, however, a ribosome itself is a ribonucleoprotein complex 
that is not an enzyme, but provides a platform where several distinct 
catalysts catalyze the respective specific reactions. For example, the 
GTPases enhance hydrolysis of GTP molecules while the peptidyl transferase 
catalyzes the formation of the peptide bond between the incoming amino 
acid monomer and the growing polypeptide.

\subsection{Comparison with some earlier works } 

Our derivation of the MM-like equation is different from the 
derivation of MM-like equation for cytoskeletal motors reported 
in ref. \cite{bustamante00} where the dwell time distribution 
was not derived. By making one-to-one correspondence beween the 
mechano-chemical cycle in their generic model for cytoskeletal 
motors and that in our model of ribosome, we find that the MM-like 
equation reported by Bustamante et al.  \cite{bustamante00} 
reduces to the MM-like equation (\ref{eq-effMM}).

In a recent work, Jackson et al.\cite{jackson08} modelled the process 
of translation as an enzymatic reaction. However, there are crucial 
differences between their formulation of translation and our 
interpretation of the mechano-chemical cycle in our model. In their 
formulation, Jackson et al.\cite{jackson08} treated the completely 
synthesized protein as the product of the enzymatic reaction, i.e., 
the run of a single ribosome from the initiation site to the 
termination site was treated as a single enzymatic reaction. In 
contrast, translocation of a ribosome from one codon to the next, 
and the associated elongation of the growing polypeptide by one 
amino acid has been treated in our calculation here as a single 
enzymatic reaction.

\section{Force-velocity relation}\label{fvrel}

\begin{figure}
\begin{center}
\includegraphics[angle=-90,width=0.8\columnwidth]{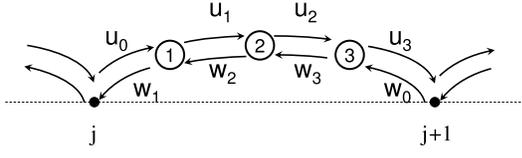}
\end{center}
\caption{The mechano-chemical cycle of the molecular motor in the 
Fisher-Kolomeisky model for $m = 4$. The horizontal dashed line 
shows the lattice which represents the track; $j$ and $j+1$ 
represent two successive binding sites of the motor. The circles 
labelled by integers denote different ``chemical'' states in 
between $j$ and $j+1$. 
}
\label{fig-fishkolo}
\end{figure}

\begin{figure} 
\begin{center}
(a)\\
\includegraphics[width=0.8\columnwidth]{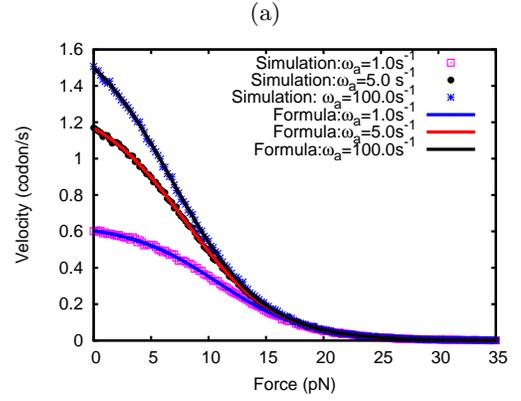}\\
\vspace{1cm}
(b)\\
\includegraphics[width=0.8\columnwidth]{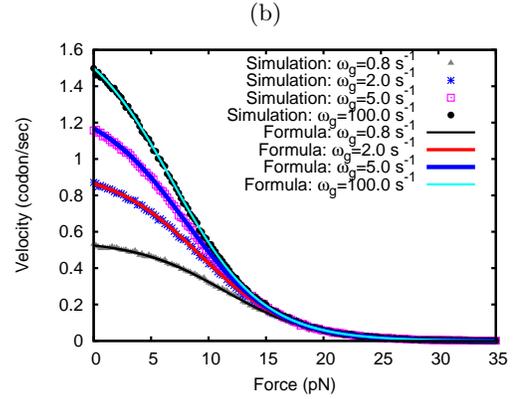}
\end{center}
\caption{(Color online) Force-velocity relation for a ribosome in our 
model for a few different values of the parameter (a) $\omega_{a}$ 
(which is proportional to the concentration of tRNA-bound amino-acid 
subunits), and (b) $\omega_{g}$ (which can be controlled by varying 
GTP concentration). The continuous curve has been obtained from the 
formula (\ref{eq-fvfinal}) whereas the discrete symbols denote the 
numerical data points obtained from computer simulations of the model. }
\label{fig-fv}
\end{figure}

Utilizing an earlier result of Derrida \cite{derrida}, Fisher and 
Kolomeisky proposed a general formula for the average velocity 
$\langle V \rangle$ of a generic model of molecular motor where 
the mechano-chemical transitions form unbranched cycles. Each 
cycle consists of $m$ intermediate ``chemical'' states in between 
the successive positions on the track of the motor 
(Fig.\ref{fig-fishkolo}). The forward transitions take place at 
rates $u_j$ whereas the backward transitions occur with the rates 
$w_j$. Choosing the unit of length to be the separation between the 
successive equispaced positions of the motor on the track, the average 
velocity $\langle V \rangle$ of the motor is given by \cite{fishkolo}
\begin{equation}
 V=\frac{1}{R_m}\biggl[1-\prod^{m-1}_{j=0}\biggl(\frac{w_j}{u_j}\biggr)\biggr] 
\label{eq-kolov}
\end{equation}
where  
\begin{equation}
R_m=\sum^{m-1}_{j=0}r_j=\sum^{m-1}_{j=0}\biggl(\frac{1}{u_j}\biggr)\biggl[1+\sum^{m-1}_{k=1}\prod^{k}_{i=1}\biggl(\frac{w_{j+i}}{u_{j+i}}\biggr)\biggr] 
\label{eq-fv3}
\end{equation}
Formally, our model of ribosome is a special case of the Fisher-Kolomeisky 
model where 
$u_0=\omega_a$, $u_1=\omega_{h1}$, $u_2=k_2$, $u_3=\omega_g$, 
$u_4=\omega_{h2}$ and $w_1=\omega_p$. Hence, in this special case 
the eqn.(\ref{eq-kolov}) can be written in a compact form as 
\begin{equation}
 V=\frac{\omega_{h2}}{1+\Omega_{h2}}
\label{eq-fvr}
\end{equation}
with  
\begin{equation}
 \Omega_{h2}=\omega_{h2}/k_{eff}
\end{equation}
and 
\begin{equation}
\frac{1}{k_{eff}} = \frac{1}{\omega_a} \biggl(1+\frac{\omega_p}{\omega_{h1}}\biggr) + \frac{1}{\omega_{h1}} + \frac{1}{k_2} + \frac{1}{\omega_g} 
\label{eq-keff}
\end{equation}
Note that $k_{eff}^{-1}$ is an effective time delay induced by the 
intermediate biochemical steps in between two successive hoppings 
of the ribosome from one codon to the next \cite{bcpre}.
Interestingly, simplification of the exact expression (\ref{mm-1}) 
yields the same formula (\ref{eq-fvr}) which we derived as a special 
case of the Fisher-Kolomeisky formula for average velocity.

In our model the load force will only affect the mechanochemical 
transition from state $5$ at $j$ to state $1$ at $j+1$. The 
dependence of the rate constant $\omega_{h2}$ on $F$ is given by
\begin{equation}
 \omega_{h2}(F)=\omega_{h2}(0)exp\biggl(-\frac{F~\delta}{k_BT}\biggr) 
\end{equation}
where $\omega_{h2}(0)$ is the magnitude of the rate constant 
$\omega_{h2}$ in the absence of load force and the typical length 
of each codon is $\delta=3 \times 0.34$nm. Thus, when subjected to a 
load force $F$, the force velocity relation for a single ribosome 
becomes
\begin{equation}
V(F) =\frac{\omega_{h2}(F)}{1+\Omega_{h2}(F)}
\label{eq-fvfinal}
\end{equation}

The force-velocity relation $\langle V \rangle(F)$ has been plotted 
in fig.\ref{fig-fv}(a) and (b) for a few different values of 
$\omega_{a}$ and $\omega_{g}$, respectively, to demonstrate the 
dependence of $\langle V \rangle(F)$ on the supply of amino acid 
monomers and the chemical fuel GTP. For fixed $\omega_{a}$ and 
$\omega_{g}$, $\langle V \rangle$ decreases with increasing $F$ and 
vanishes at $F = F_{s}$ which is identified as the corresponding 
{\it stall force}. Moreover, for a given $F$, $\langle V \rangle$ 
increases monotonically with increasing $\omega_{a}$ and $\omega_{g}$ 
although the rate of increase gradually slows down. It is interesting 
to note that $F_{s}$ is independent of both $\omega_{a}$ and $\omega_{g}$ 
because, at stall, a ribosome uses neither amino acid monomers nor GTP. 
For the typical values of the rate constants, which we have used in 
fig.\ref{fig-fv}, $F_{s} \simeq 25-27$ pN. This theoretical estimate 
is consistent with the value $26.5$ pN reported by Sinha et al. 
\cite{sinha04}.

\section{Fluctuations: mean square dwell time and diffusion constant}\label{fluc}

\begin{figure}
\begin{center}
(a)\\
\includegraphics[width=0.8\columnwidth]{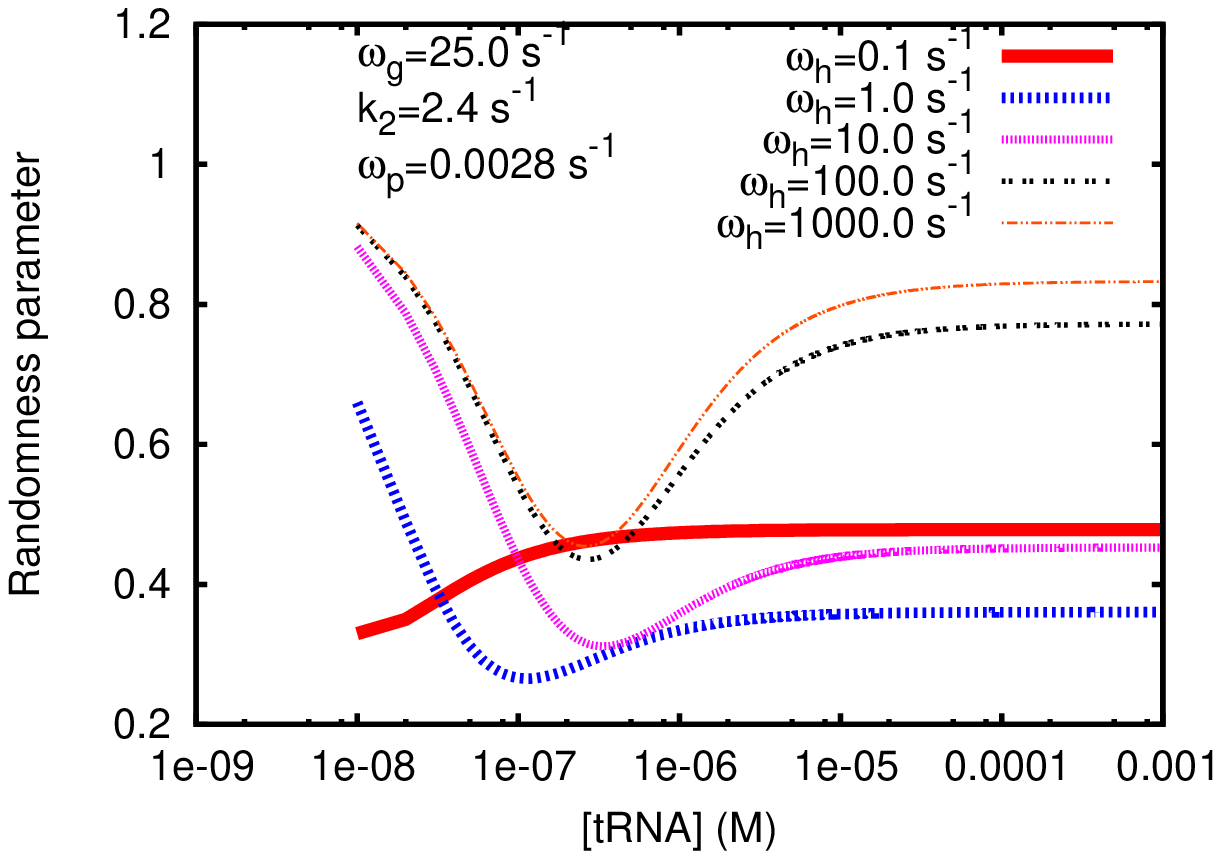} 
\vspace{1cm}\\
(b)\\
\includegraphics[width=0.8\columnwidth]{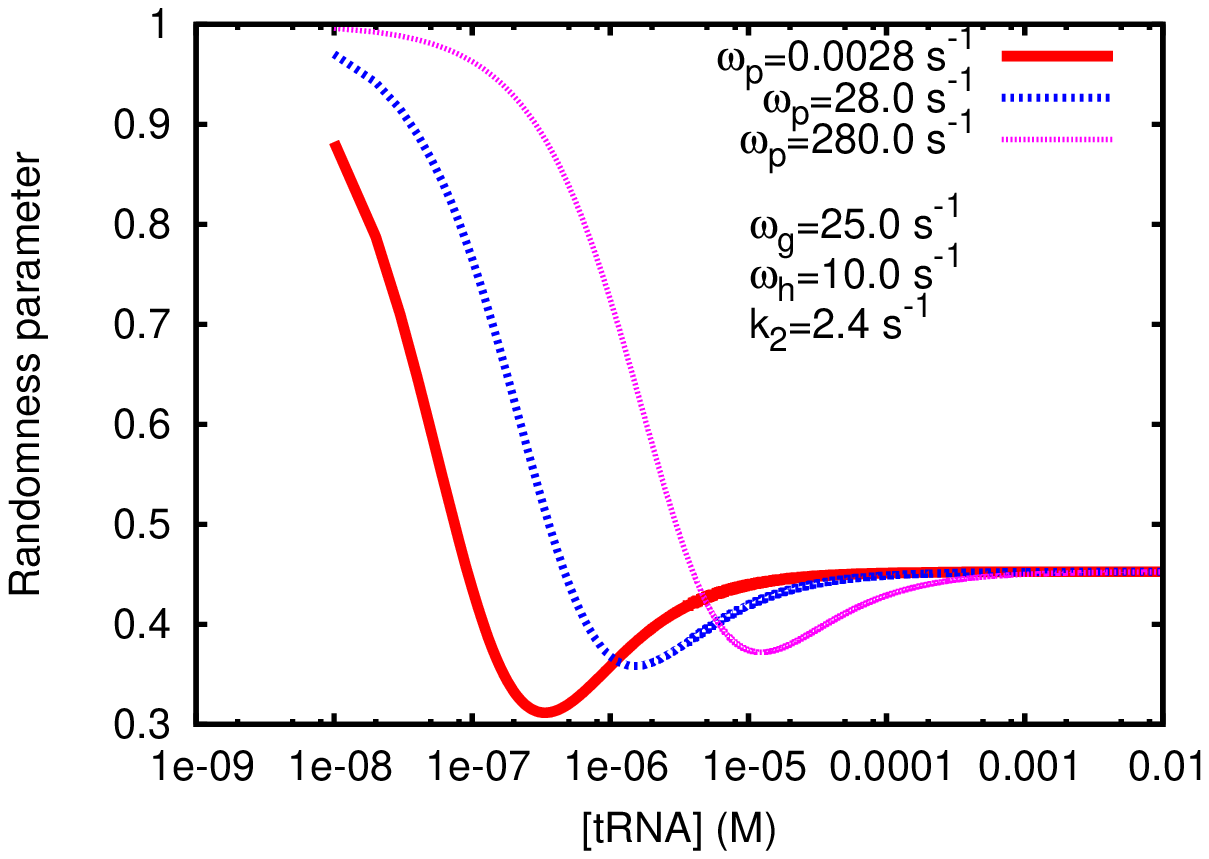}
\end{center}
\caption{(Color onlne) The randomness parameter $r$, defined by the equation 
(\ref{eq-ranpar}), is plotted against the concentration of 
aa-tRNA for a few different values of $\omega_{h}$ (in (a)) and 
$\omega_{p}$ (in (b)).}
\label{fig-ranpar}
\end{figure}

\begin{figure}
\begin{center}
\includegraphics[width=0.8\columnwidth]{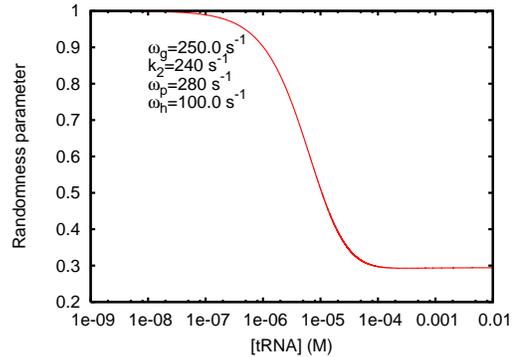}
\end{center}
\caption{(Color online) The randomness parameter $r$, defined by 
the equation (\ref{eq-ranpar}), is plotted against the concentration 
of aa-tRNA for a set of parameter values where all the magnitudes of 
all the rate constants, other than $\omega_{a}$, are quite high.} 
\label{fig-ranpar2}
\end{figure}

\subsection{Fluctuations in dwell times}

Mean-square dwell time is defined by 
\begin{equation}
 \langle t^2 \rangle = \int^{\infty}_{0}t^2f(t)dt.
\end{equation}
For our model 
\begin{equation}
\langle t^2 \rangle = 2\biggl[\frac{C_1}{\omega_1^3}+\frac{C_2}{\omega_2^3}+\frac{C_3}{k^3_2}+\frac{C_4}{\omega^3_g}+\frac{C_5}{\omega^3_{h2}}\biggr]. 
\label{eq-msqdwell} 
\end{equation}
The expression (\ref{eq-msqdwell}) can be expressed also as 
\begin{equation}
\langle t^2 \rangle = 2(\langle t \rangle^2-\xi_2)
\label{eq-msqtgen}
\end{equation}
where,
\begin{eqnarray}
\xi_{2} &=& \biggl(\frac{\omega_p}{\omega_a~\omega_{h1}}\biggr)\biggl(\frac{1}{k_2}+\frac{1}{\omega_g}+\frac{1}{\omega_{h2}}\biggr)\nonumber \\ 
&+& \frac{1}{\omega_a \omega_{h1}}+\frac{1}{\omega_a k_2}+\frac{1}{\omega_a \omega_g}+\frac{1}{\omega_a \omega_{h2}}\nonumber \\ 
&+& \frac{1}{\omega_{h1} k_2}+\frac{1}{\omega_{h1}\omega_g}+\frac{1}{\omega_{h1}\omega_{h2}}\nonumber \\ 
&+& \frac{1}{k_2 \omega_g}+\frac{1}{k_2 \omega_{h2}}+\frac{1}{\omega_g \omega_{h2}} 
\end{eqnarray}
Note that only the first term involves $\omega_{p}$. The remaining ten 
terms are inverse of the products of the five rate constants.

Let us define ``randomness parameter'' $r$ as
\begin{equation}
r=\frac{\left\langle  t^2 \right\rangle - {\left\langle t \right\rangle}^2}{{\left\langle t \right\rangle}^2}
\label{eq-ranpar}
\end{equation}
Note that $r$ is a quantitative measure of the fluctuations in the 
dwell times of a ribosome. By substiuting the expressions of 
$\left\langle t^2 \right\rangle$ and $\left\langle t \right\rangle$ 
into (\ref{eq-ranpar}) we obtain,
\begin{equation}
r=\frac{\langle t \rangle^2-2\xi_2}{\langle t \rangle^2}
\label{eq-randribo}
\end{equation}
A non-trivial feature of the expression (\ref{eq-randribo}) is that 
it cannot be obtained simply by substituting $\omega_{-1}^{eff}$ 
and $\omega_{2}^{eff}$ into the expression for $r$ derived by Kou 
et al.\cite{kou05} for the two-step Michaelis-Menten enzymatic reaction. 
In other words, the fluctuations of the dwell times in the five-step 
model for the kinetics of ribosomes cannot be captured by the effective 
two-state model drawn in fig.\ref{fig-riboeffMM}.

The randomness parameter $r$ is plotted in fig.\ref{fig-ranpar} 
as a function of the tRNA concentration for a few different 
values of the parameter $\omega_{h}$ (in (a)) and $\omega_{p}$ 
(in (b)). We find (not shown in any figure) that the numerator of 
$r$ (i.e., $\langle t^2 \rangle - \langle t \rangle^2$) decreases 
monotonically with increasing concentration of $tRNA$; it is the 
variation of the denominator of $r$ with $tRNA$ concentration 
that is responsible for the non-monotonic variation of $r$.

It is well known \cite{kou05} that, for a one-step Poisson process, 
$r = 1$. At extremely low concentrations of aa-tRNA, the binding of 
a correct species of aa-tRNA to the A site on the large subunit of a 
ribosome is the rate-limiting step in its mechano-chemical cycle. 
Therefore, $r$ is unity at sufficiently low values of aa-tRNA. $r$ 
decreases with the increase of aa-tRNA concentration. This decrease 
is caused by  the formation of intermediate complexes which also 
affect the rates of progress of the mechano-chemical cycle. However, 
with the further increase of aa-tRNA concentration, the randomness 
parameter $r$ increases again. Finally, the randomness parameter 
saturates to a value which is determined by the number of rate-limiting 
steps in the mechano-chemical cycle. Such nonmonotonic variation of 
$r$ with aa-tRNA concentration reduces to a monotonic decrease when the 
magnitudes of the rate constants are sufficiently high (see 
fig.\ref{fig-ranpar2}).

\subsection{Diffusion constant}

The diffusion constant $D$ is a measure of fluctuations around the 
directed movement of the ribosome, on the average, in space. We 
now derive a closed form expression for $D$ and relate it to the 
fluctuations in the dwell times. Fisher and Kolomeisky's general 
result for diffusion coefficient $D$ is 
\begin{equation}
 D=\biggl[ \frac{(VS_N+dU_N)}{R^2_N}-\frac{(N+2)V}{2} \biggr]\frac{d}{N}
\end{equation}
where \begin{equation}
 S_N=\sum^{N-1}_{j=0}s_j\sum^{N-1}_{k=0}(k+1)r_{k+j+1}
\end{equation}
and
\begin{equation}
 U_N=\sum^{N-1}_{j=0}u_jr_js_j
\end{equation}
while,
\begin{equation}
 s_j=\frac{1}{u_j}\biggl(1+\sum_{k=1}^{N-1}\prod^{k}_{i=1}\frac{w_{j+1-i}}{u_{j-i}}\biggr)
\end{equation}
\begin{equation}
 R_N=\sum^{N-1}_{j=0}r_j
\end{equation}
In our units $d=1$. Therefore, in our model the expression for $D$ becomes, 
\begin{equation}
D = (\langle t \rangle^2-2\xi_2)/{2 \langle t \rangle^3}
\end{equation}
Finally, we observe that $r$, which is a measure of the fluctuations 
in the dwell times, is related to $D$ and $\langle V \rangle$ by 
\cite{kolofish}. 
\begin{equation}
r=\frac{2D}{\langle V \rangle}
\end{equation}

\section{Distribution of run times}\label{run}

\begin{figure}
\begin{center} 
 \includegraphics[width=0.8\columnwidth]{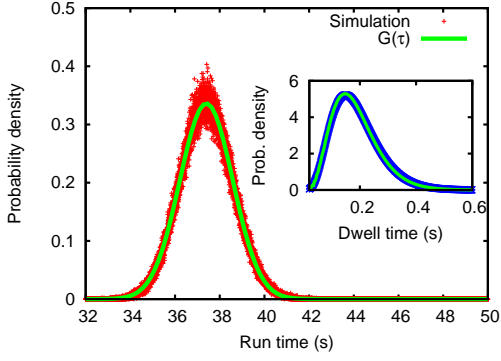} \\
\end{center}
\caption{(Color online) Distribution of the run times of a ribosome in 
our model. The continuous curve is the Gaussian distribution predicted 
by our theory while the discrete data points have been obtained from 
computer simulations. The inset shows the dwell time distribution for 
the same set of parameter values.
}
\label{fig-travtime}
\end{figure}

In this section we report the distribution of the run times $\tau$ 
of an individual from the start codon to the stop codon. The run  
time is related to the dwell times by the relation 
\begin{equation}
\tau = \sum_{j=1}^{L} t_{j} 
\end{equation}
Central limit theorem states that, as $L \to \infty$, the distribution 
$G(\tau)$ of the run times $\tau$ approaches a Gaussian, irrespective 
of the nature of the distribution of the dwell times, since the dwell 
times at different codons are independent of each other. Obviously, 
for sufficiently large $L$ \cite{vankampen}, 
\begin{equation}
G(\tau) = \frac{1}{\sqrt(2\pi\sigma_{\tau}^2)} exp(-\frac{(\tau-\langle \tau \rangle)^2}{2\sigma_{\tau}^2})
\end{equation}
where
\begin{equation}
\langle \tau \rangle = L \langle t \rangle 
\end{equation}
and
\begin{equation}
\langle \tau^{2} \rangle - \langle \tau \rangle^2 = L (\langle t^{2} \rangle - \langle t \rangle^2).
\end{equation}
Using $\langle t \rangle$ from (\ref{mm-1}) and $\langle t^{2} \rangle$ 
from (\ref{eq-msqtgen}), we obtain the Gaussian distribution $G(\tau)$.
The gaussian distribution $G(\tau)$ thus obtained is plotted in 
fig.\ref{fig-travtime}; it is in excellent agreement with the 
corresponding numerical data obtained from computer simulations.

\section{Effects of steric interactions of ribosomes}\label{interaction}

The average velocity of a ribosome is also the mean rate of polymerization 
of a protein. We define the {\it flux} of ribosomes to be the total number 
of ribosomes leaving the stop codon (i.e., $j = L$) per unit time. 
Obviously, the overall rate of protein synthesized from a single mRNA 
template is identical to the flux of the ribosomes on that mRNA track. 
The {\it number density} of the ribosomes is given by $\rho=N/L$. 
The size of a typical ribosome is such that, simultaneously, it covers 
${\ell}$ codons where ${\ell} \gg 1$. We treat ${\ell}$ as a parameter 
of the model. For a given number $N$ of ribosomes, the total fraction 
of the lattice covered by all the ribosomes is given by the {\it coverage 
density} $\rho_{cov}=N{\ell}/L$. 

In the preceeding sections, we have ignored the possibility of steric 
interactions among the ribosomes. Consequently, the average velocity 
was independent of the ribosome population on the given mRNA track. 
Such a scenario holds at most at sufficiently low coverage densities. 
However, in the presence of inter-ribosome interactions the average 
velocity becomes a function of the coverage density thereby giving 
rise to non-trivial variation of the flux $J$ (and, hence, the overall 
rate of protein synthesis) with $\rho_{cov}$. Moreover, the density 
profile of the ribosomes on the track also exhibits interesting features.
In this section we study the spatio-temporal organization of the ribosomes 
in terms of the flux as well as the density profiles on a single mRNA 
track and plot the phase diagrams of the model. 

Let $P_{\mu}(j,t)$ be the probability of finding a ribosome at site $j$, 
in the chemical state $\mu$ at time $t$. Also, 
$P(j,t) = \sum_{\mu=1}^{5} P_{\mu}(j,t)$, is the probability of finding
a ribosome at site $j$, irrespective of its chemical state. Let 
$P(\underbar{j}|k)$ be the conditional probability that, given a ribosome 
at site $j$, there is another ribosome at site $k$. Then,
$Q(\underbar{j}|k) = 1 - P(\underbar{j}|k)$ is the conditional
probability that, given a ribosome in site $j$, site $k$ is empty. In
the mean-field approximation, the Master equations for the
probabilities $P_{\mu}(j,t)$ are given by \cite{bcpre}
\begin{eqnarray} 
\frac{\partial{}P_{1}(j,t)}{\partial{}t} &=& \omega_{h2} P_{5}(j-1,t) Q(\underbar{j-1}|j-1+\ell) \nonumber \\ 
&+& \omega_{p} P_2(j,t) - \omega_{a} P_{1}(j,t)
\label{eq-mint1}
\end{eqnarray}
\begin{equation} 
\frac{\partial{}P_{2}(j,t)}{\partial{}t} = \omega_{a} P_{1}(j,t) - (\omega_{p} + \omega_{h1}) P_{2}(j,t)
\label{eq-mint2}
\end{equation}
\begin{equation} 
\frac{\partial{}P_{3}(j,t)}{\partial{}t} = \omega_{h1} P_{2}(j,t) - k_{2} P_{3}(j,t)
\label{eq-mint3}
\end{equation}
\begin{equation} 
\frac{\partial{}P_{4}(j,t)}{\partial{}t} = k_{2} P_{3}(j,t) - \omega_{g} P_{4}(j,t)
\label{eq-mint4}
\end{equation}
\begin{equation} 
\frac{\partial{}P_{5}(j,t)}{\partial{}t} = \omega_{g} P_{4}(j,t) - \omega_{h2} P_{5}(j,t) Q(\underbar{j}|j+{\ell}) 
\label{eq-mint5}
\end{equation}
Because of the normalization condition 
\begin{equation}\label {gg}
P(j,t)=\sum_{\mu=1}^{5}P_{\mu}(j,t)=\frac{N}{L} = \rho
\end{equation}
not all of the five $P_{\mu}(j,t)$ are independent. 

\subsection{Effects of steric interactions on rate of protein synthesis}\label{flux}

The dwell time distribution $f(t)$ certainly gets affected by the steric 
interactions. As a first step, we have calculated the effects of the 
interactions on the average velocity which is just the inverse of the 
mean dwell time. 

Under periodic boundary conditions, in the steady state, $P_{\mu}(j,t)$ 
become independent of $j$ and $t$. From the steady-state limit of the 
equations (\ref{eq-mint1})-(\ref{eq-mint5}), we derived the expressions 
$P_{\mu}$ ($\mu=1,2,...5$) and the flux \cite{bcpre}
\begin{equation}
 J_{PBC}= \rho \langle V \rangle = \frac{\omega_{h2}\rho(1-\rho l)}{(1+\rho-\rho l)+\Omega_{h2}(1-\rho l)}
\label{eq-pbcj}
\end{equation}
where 
\begin{equation}
\Omega_{h2} = \omega_{h2}/k_{eff}.
\end{equation}
and $k_{eff}$ is given by (\ref{eq-keff}).

From (\ref{eq-pbcj}) the $\rho$-dependent average velocity 
$\langle V \rangle$ can be obtained by dividing $J$ by $\rho$. At 
sufficiently low densities this expression reduces to
\begin{equation}
J=\frac{\omega_{h2}\rho}{(1+\Omega_{h2})}
\end{equation}
and, hence, we recover the exact formula (\ref{eq-fvr}) for the 
average velocity of a single ribosome.

\subsection{Phase diagrams under open boundary conditions}\label{phase}

\begin{figure}
 \begin{center} 
(a)\\
\includegraphics[width=0.9\columnwidth]{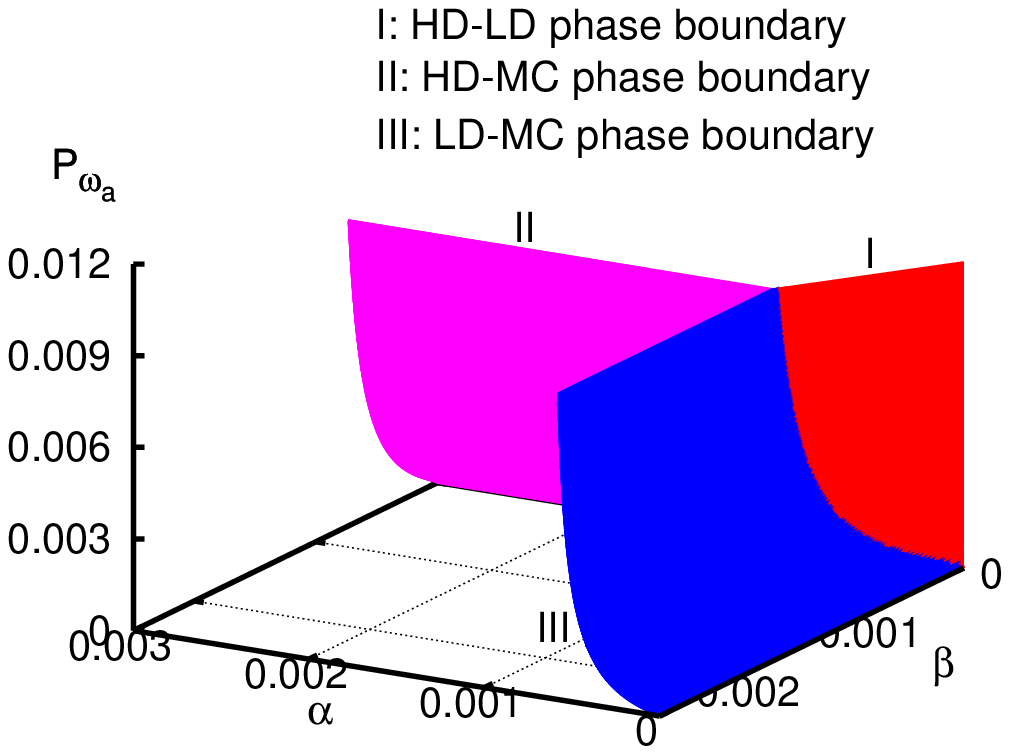}\\
\vspace{1cm}
(b)\\
\includegraphics[width=0.9\columnwidth]{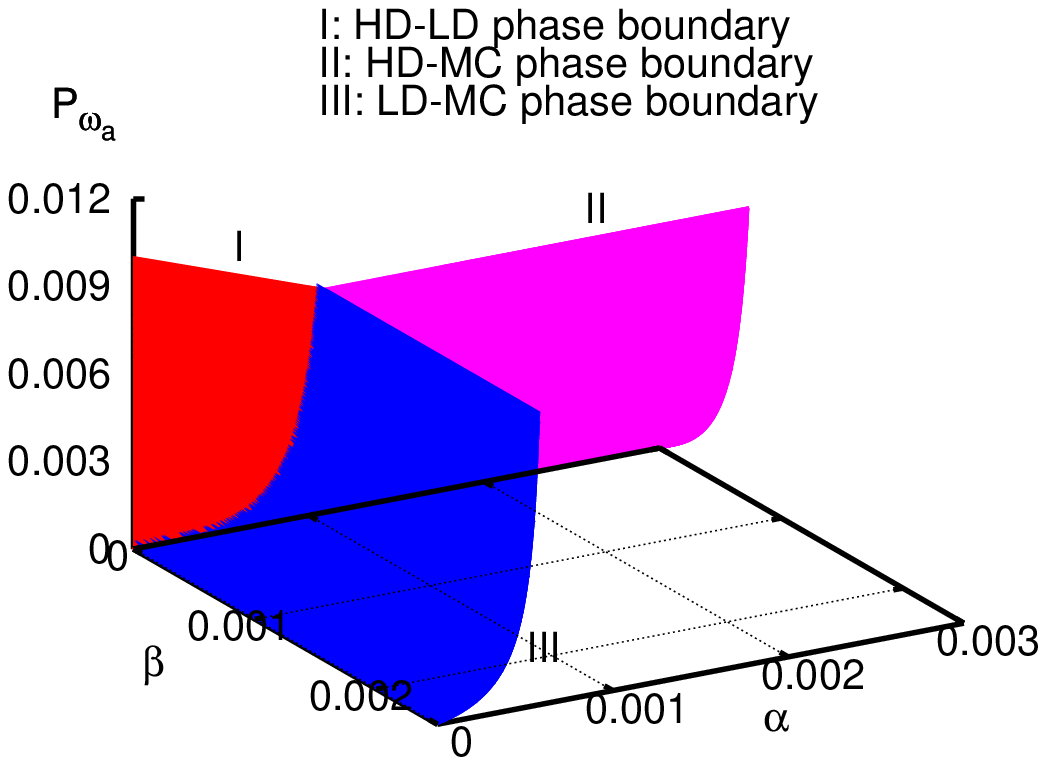}
\end{center}
\caption{(Color online) A 3d-phase diagram of our ribosome traffic model. 
The LD and HD phases coexist on the surface I (colour red). Surfaces II 
(colour purple) and III (colour blue) seperate the MC phase from the HD 
and LD phases, respectively. Phase diagram is shown in (a) and (b) 
from two different orientations. The parameters used are 
$\omega_p=0.0028 s^{-1}, \omega_a=25.0, k_2=2.4 s^{-1}, \omega_g=25.0 s^{-1}$}
\label{fig-3d1}
\end{figure} 

\begin{figure}
 \begin{center} 
(a)\\
\includegraphics[width=0.9\columnwidth]{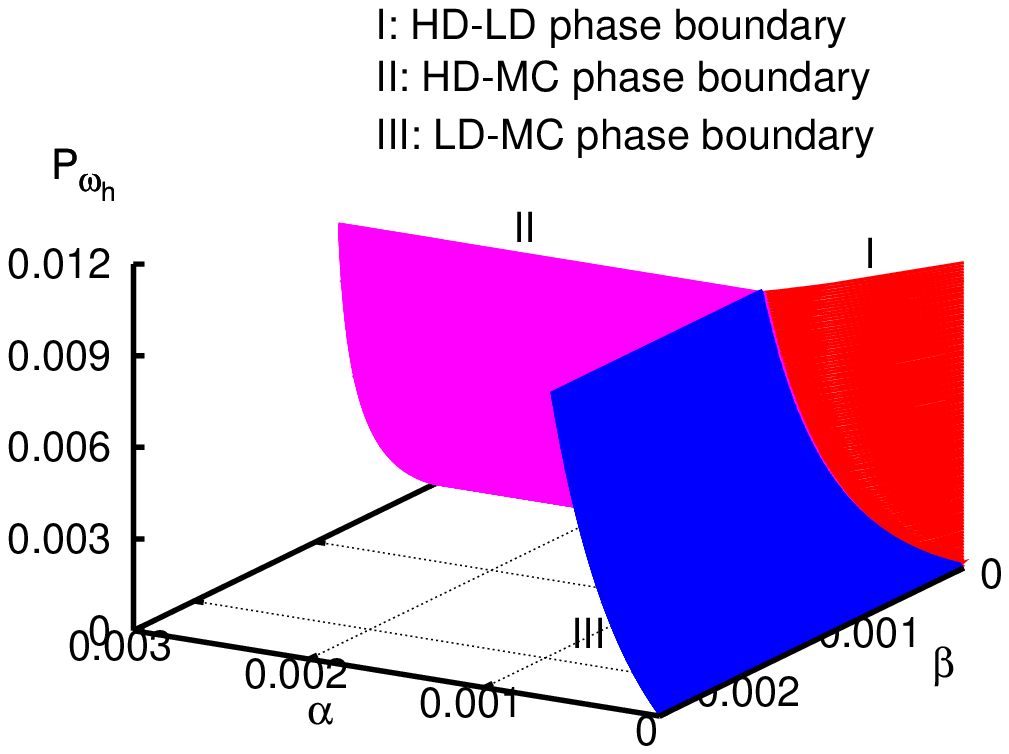}\\
\vspace{1cm}
(b)\\
\includegraphics[width=0.9\columnwidth]{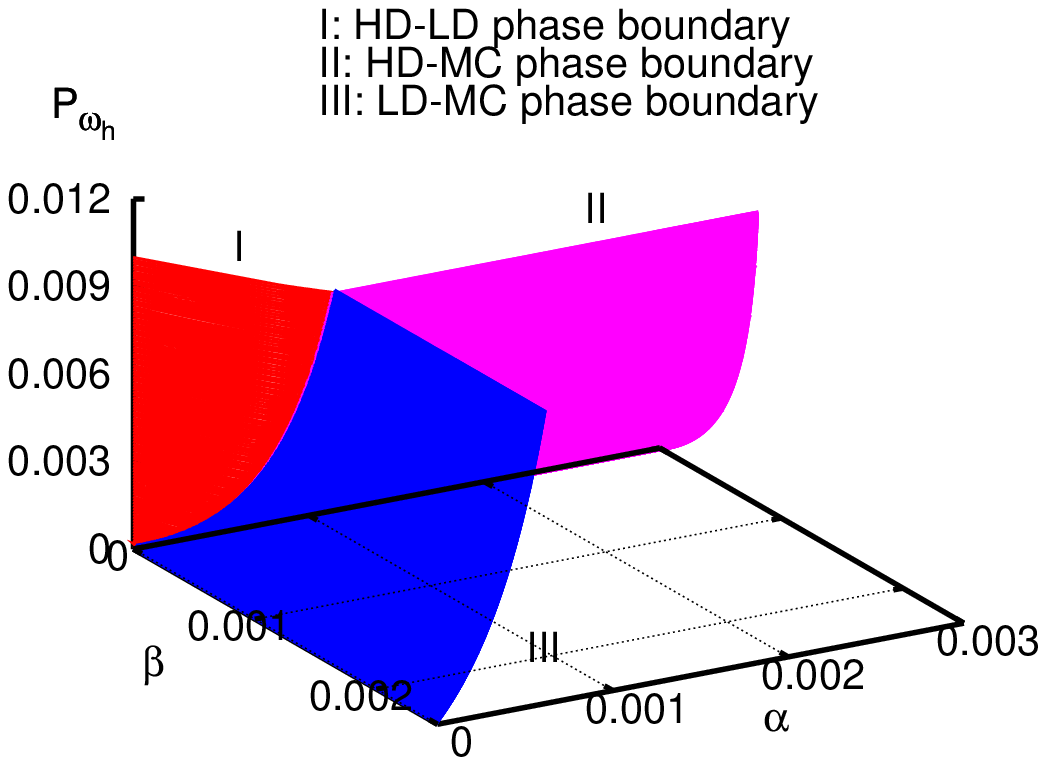}
\end{center}
\caption{(Color online) Same as in fig.\ref{fig-3d1} except that 
$\omega_{h}$ has been used instead of $\omega_{a}$.  }
\label{fig-3d2}
\end{figure}

\begin{figure}
 \begin{center} 
(a)\\
\includegraphics[width=0.9\columnwidth]{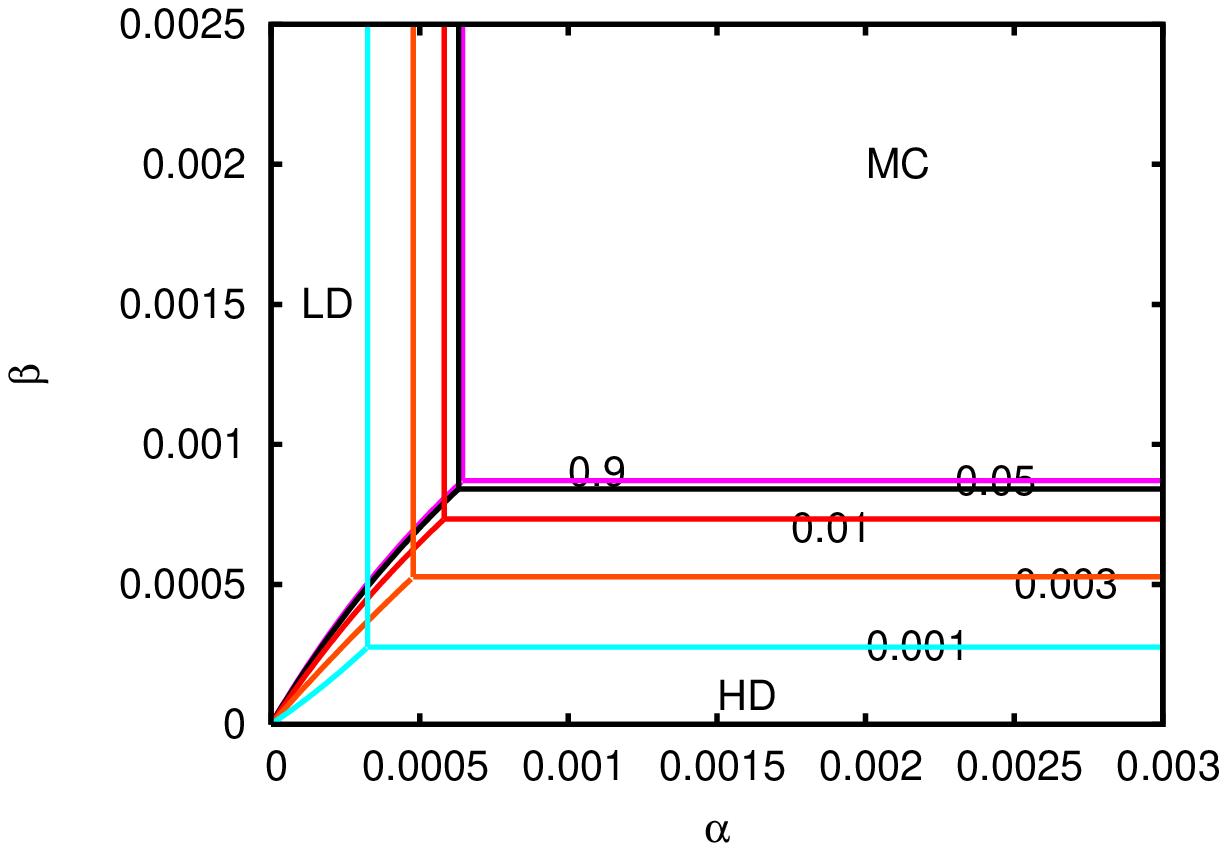}\\
\vspace{1cm}
(b)\\
\includegraphics[width=0.9\columnwidth]{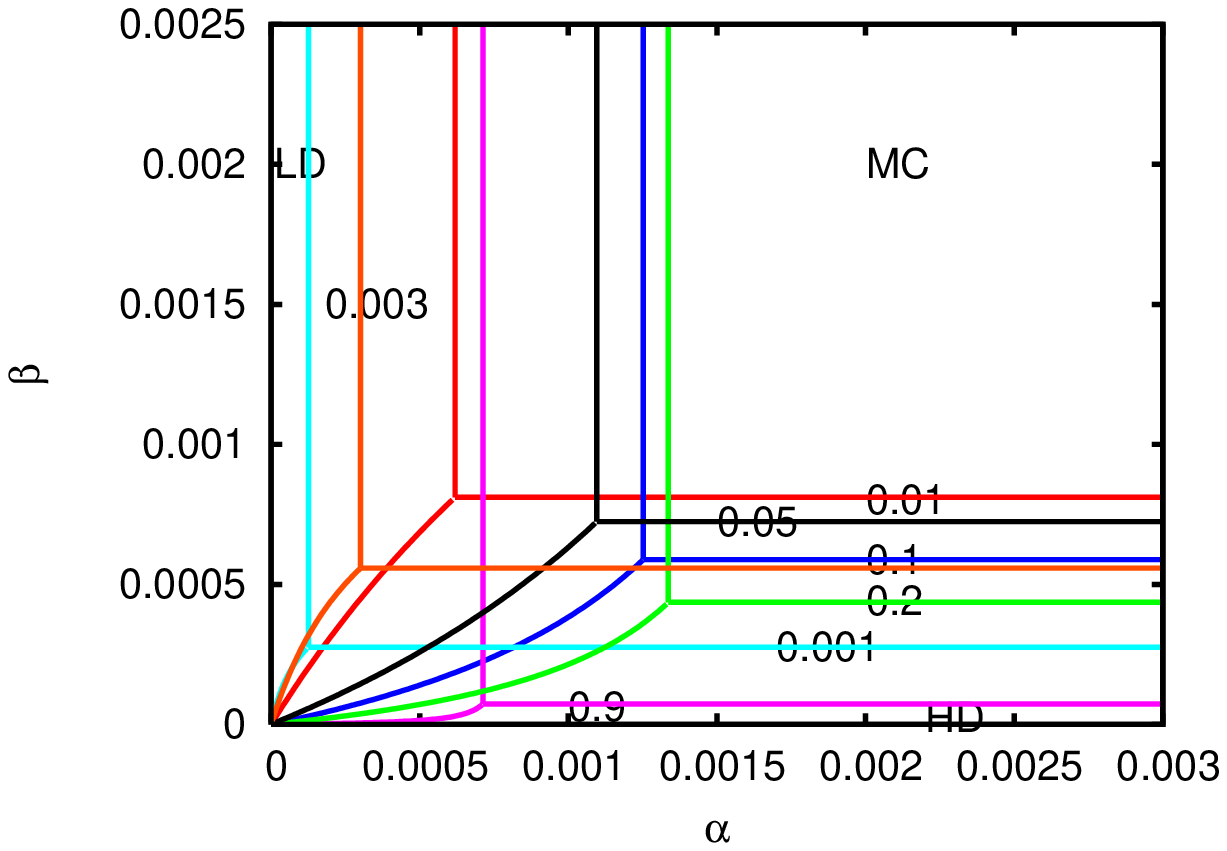}
\end{center}
\caption{(Color online) 2d-cross sections of the 3d phase diagrams of 
the ribosome traffic model parallel to the $\alpha$-$\beta$ plane for 
different values of (a) $P_{{\omega}_a}$, and (b) $P_{{\omega}_h}$, 
all projected onto the same figure. The numbers on the phase boundary 
lines represent different values of (a) $P_{{\omega}_a}$ and 
(b) $P_{{\omega}_h}$. The HD and LD phases coexist on the curved lines 
whereas the straight lines separate the MC phase from the HD and LD 
phases, as shown.  The parameters used for this figure are 
$\omega_p=0.0028 s^{-1},\omega_a=25.0, k_2=2.4 s^{-1}, \omega_g=25.0 s^{-1}$}
\label{fig-2d}
\end{figure}

Initiation and termination of protein synthesis are captured more 
realistically by imposing open boundary conditions (OBC) than by 
the periodic boundary conditions (PBC). Whenever the first
${\ell}$ sites on the mRNA are vacant this cluster of sites is allowed
to be covered by a fresh ribosome with the probability $\alpha$ in
the time interval $\Delta t$ (in all our numerical calculations we
take $\Delta t = 0.001$ s). Since $\alpha$ is the probability of 
initiation in time $\Delta t$, the corresponding rate constant (i.e., 
probability of initiation per unit time) $\omega_{\alpha}$ is related 
to $\alpha$ by $\alpha{}=1-e^{-\omega_{\alpha}\times{}\Delta t}$. 
Similarly, whenever the rightmost ${\ell}$ sites of the mRNA lattice 
are covered by a ribosome, i.e., the ribosome is bound to the $L$-th 
codon, the ribosome gets detached from the mRNA with probability 
$\beta$ in the time interval $\Delta t$; the corresponding rate 
constant being denoted by $\omega_{\beta}$. For all further discussions 
in this paper, we'll assume $\omega_{h1} = \omega_{h2} = \omega_{h}$ 
because both these processes are driven by GTP hydrolysis.

TASEP is known to exhibit three dynamical phases, namely, high-density 
(HD) phase, low-density (LD) phase and the maximal current (MC) phase 
in the $\alpha-\beta$ plane. Our main interest is to explore the nature 
of the dynamical phases in different regions of the 4-dimensional space 
spanned by $\alpha$, $\beta$, $P_{\omega_a}$, $P_{\omega_h}$.

The parameters $\omega_{a}$ and $\omega_{h}$ can be controlled by
varying the concentrations of the aa-tRNA molecules and GTP
molecules in the solution.  The parameter $\alpha$ 
is determined by the rate of assembling of the large and small 
subunits of a ribosome, their final coupling on the initiation 
site and the assistance of several other regulatory proteins in 
the initiation of the actual polymerization of a protein. 
Strictly speaking, a single parameter $\beta$ captures essentially 
two different events both of which take place at the termination 
site $j=L$. After the full protein has been polymerized, the 
ribosome releases the protein into the surrounding medium and then 
dissociates from the mRNA track (the decoupling of the two subunits 
also takes place; these are then recycled for another round of 
protein synthesis) \cite{hirokawa06,petry08}. Therefore, the value 
$\beta = 1$, which we assumed in ref.\cite{bcpre} is, in general, 
not very realistic. Even in the special case $\beta=1$, in 
ref.\cite{bcpre} we reported only a couple of two-dimensional cross 
sections of the full phase diagram of this model. In this paper we 
plot phase diagrams in three-dimensional spaces spanned by 
$\alpha-\beta-P_{\omega_a}$ and $\alpha-\beta-P_{\omega_h}$.

For plotting the phase diagram, we use the same extremum principle
\cite{krug91,popkov2,hager1,hager2} which we used in ref.\cite{bcpre}.
In this approach, we imagine that the left and right boundaries of the
system are connected to two reservoirs with particle densities 
$\rho_{-}$ and $\rho_{+}$, respectively. These two reservoirs are 
essentially two infinite lattices with the number densities $\rho_{-}$ 
and $\rho_{+}$, respectively. We calculate the unknown densities 
$\rho_{-}$ and $\rho_{+}$, in terms of the rate constants of our 
model, by imposing the requirement that these reservoirs give rise to 
the same probabilities $\alpha$ and $\beta$, of hopping with which a 
ribosome enters and exits, respectively, the open system. 

The extremum principle then relates the flux ${\cal J}$ in the open
system to the flux $J(\rho)$ for the corresponding closed system
(i.e., the system with periodic boundary conditions). Extremum current 
hypothesis \cite{krug91,popkov2,hager1,hager2} states that, for the 
open system connected to the two reservoirs of number densities 
$\rho_{-}$ and $\rho_{+}$ at its entrance and exit, the flux ${\cal J}$ 
is related to the corresponding flux $J_{PBC}$ in the closed system by  
\begin{eqnarray}
{\cal J}  = \left\{ \begin{array}{ll}
         \mbox{max} ~~ J_{PBC}(\rho) & \mbox{if}~~ \rho_{-}~ > ~\rho~ > ~\rho_{+}\\
         \mbox{min} ~~ J_{PBC}(\rho)& \mbox{if}~~ \rho_{-}~ < ~\rho~ < ~\rho_{+}.\end{array} \right. \nonumber\\
\end{eqnarray}
Since the flux-density relation (also called the fundamental diagram) 
of our model of ribosome traffic under periodic boundary conditions 
exhibits a single maximum, the extremum principle reduces to a simpler 
maximum current principle (MCP). According to this MCP, in the limit 
$L\to{}\infty$,
\begin{eqnarray}
{\cal J}  = \mbox{max} ~~ J_{PBC}(\rho) & \mbox{if}~~ \rho_{-}~ > ~\rho~ > ~\rho_{+}.
\end{eqnarray}

\subsubsection{Calculation of $\rho_{*}, \rho_{-}$ and $\rho_{+}$}

From (\ref{eq-pbcj}), the maximum flux under PBC corresponds to the 
number density 
\begin{eqnarray}
\rho_{*} = \sqrt{\biggl(\frac{1+\Omega_{h2}}{\ell}\biggr)} \biggl[\frac{1}{\sqrt{{\ell}(1+\Omega_{h2})}+1}\biggr]
\label{soln2}
\end{eqnarray}

Next we calculate $\rho_{-}$.
We use symbol $1$ to represent the sites covered by ribosome while the 
symbol $0$ represents the sites which are not covered by any ribosome. 
Let $P^{jump}_{-}$ be the probability that, given an empty site, from 
left a ribosome will hop onto it in the next time step. We have
\begin{equation}  
P^{jump}_{-} = P(\underbrace{1..............1}_{\ell}|\underline{0})\times P_5 \times \omega_{h2}\times \Delta~t
\end{equation}
where $P_5$ is the probability of finding ribosome in state $5$ inside 
the reservoir and the conditional probability 
$P(\underbrace{1..............1}_{\ell}|\underline{0})$ represents that, 
given an empty site, there will be a ribosome in the adjacent $\ell$ 
sites to its left. 
On the basis of arguments similar to those presented in ref.\cite{bcpre}, 
we get  
\begin{equation}
P(\underbrace{1..............1}_{\ell}|\underline{0})=\frac{\rho}{(1+\rho-\rho\ell)}
\end{equation}
Moreover, as argued in ref.\cite{bcpre}, 
\begin{equation}
P_5=\frac{1}{1+\Omega_{h2}}
\end{equation}
Now, $\rho_{-}$ is the solution of the equation $\alpha=P^{jump}_{-}$ 
and, hence, we get,
\begin{equation}
 \rho_{-}=\frac{\alpha(1+\Omega_{h2})}{\alpha(1+\Omega_{h2})(\ell -1)+P_{\omega_{h2}}}
\label{rm}
\end{equation}
where $P_{\omega_{h2}}$ is the probability of hydrolysis in time $\Delta~t$.

Following similar arguments, we now calculate $\rho_{+}$. The  
probability $P^{jump}_{+}$, given that a ribosome which covers $\ell$ 
successive sites, will hop onto the next adjacent empty site to its 
right in the next time step,
\begin{equation}
P^{jump}_{+}=P(\overbrace{\underline{1..............1}}^{\ell}|~0)\times P_5\times P_{\omega_{h2}}
\end{equation}
where $P(\overbrace{\underline{1..............1}}^{\ell}|~0)$ is the 
conditional probability of finding a hole at site $j$, given that the site 
$(j-\ell-1)$ is occupied by the leftmost part of the ribosome. It is 
straightforward to show that
\begin{equation}
P(\overbrace{\underline{1..............1}}^{\ell}|~0)=\frac{1-\rho\ell}{1+\rho-\rho\ell}
\end{equation}
Now, $\rho_{+}$ is the solution of the equation $\beta=P^{jump}_{+}$; 
and, hence, we get
\begin{equation}
 \rho_{+}=\frac{\beta(1+\Omega_{h2})-P_{\omega_{h2}}}{\beta(1+\Omega_{h2})(\ell-1)-\ell P_{{\omega}_{h2}}}
\label{rp}
\end{equation}
In the limit $k_{eff} \to \infty$, the expressions (\ref{rp}) and 
(\ref{rm}) for $\rho_{+}$ and $\rho_{-}$ reduce to the corresponding 
expressions $\rho_{-} = \alpha$ and $\rho_{+} = 1-\beta$, respectively, 
for TASEP.

\subsubsection{Surface separating LD and MC phases}

The MCP imposes the condition
\begin{equation}
\rho_{-} = \rho_{*}
\label{eq-ldmcsurf}
\end{equation} 
on the surface which separates the LD and MC phases. Substituting the 
expressions (\ref{rm}) for $\rho_{-}$ into equation (\ref{eq-ldmcsurf}) 
we obtain 
\begin{equation}
 \alpha=\frac{P_{\omega_{h2}}\rho_{*}}{(1+\Omega_{h2})(1-\rho_{*}(\ell-1))}
\label{eq-ldmcalpha}
\end{equation}

\subsubsection{Surface separating HD and MC phases}

From the MCP 
\begin{equation}
\rho_{*}=\rho_{+} 
\label{eq-hdmcsurf}
\end{equation}
on the boundary between the HD and MC phases. Using the expression 
(\ref{rp}) for $\rho_{+}$, we obtain 
\begin{equation}
\beta=\frac{P_{\omega_{h2}}(1-\rho_{*}~\ell)}{(1+\Omega_{h2})(1-\rho_{*}(\ell -1))} 
\label{eq-hdmcbeta}
\end{equation} 

\subsubsection{Surface of co-existence of HD and LD phases}

Since the HD and LD phases co-exist on the surface separating these two 
phases, we obtain the boundary by solving the equation
\begin{equation}
J_{PBC}(\rho_{-}) = J_{PBC}(\rho_{+}) 
\label{eq-ldhd} 
\end{equation}
because the same current passes through the two coexisting phases in
the steady state, where the density on the entry side is $\rho_{-}$
and that on the exit side is $\rho_{+}$.

Now incorporating the expressions of $\rho_{-}$ from eqn. (\ref{rm}) 
and $\rho_{+}$ from eqn. (\ref{rp}) into eqn.(\ref{eq-pbcj}) and using 
eqn. (\ref{eq-ldhd}) we find 
that the equation of the surface of coexistence of LD and HD phases 
is given by 
\begin{equation}
 \alpha=\frac{P_{\omega_{h2}}\beta(1+\Omega_{h2})}{P_{\omega_{h2}}\ell+\beta(1-\ell+2\Omega_{h2}-\ell \Omega_{h2}+\Omega^2_{h2})}
\end{equation}
or, equivalently, 
\begin{equation}
\beta=\frac{\alpha~\ell~P_{\omega_{h2}}}{(1+\Omega_{h2})(P_{\omega_{h2}}-\alpha+\ell \alpha-\Omega_{h2}\alpha)}
\end{equation}

\subsubsection{Phase diagrams}

In fig.\ref{fig-3d1} we have plotted a 3d phase diagram of the ribosome 
traffic model, in the $\alpha-\beta-P_{\omega_a}$ space, which we obtained 
by following the MCH-based approach explained above. The corresponding 
3d phase diagram in the  $\alpha-\beta-P_{\omega_{h2}}$ space in plotted 
in fig.\ref{fig-3d2}. The LD and HD phases coexist on the surface I. A 
first order phase transition takes place across this surface. Surfaces  
II and III seperate the MC phase from the HD and LD phases, respectively. 
The 3d phase diagrams plotted in figs.\ref{fig-3d1}(a) and \ref{fig-3d2}(a) 
are differently oriented in figs.\ref{fig-3d1}(b) and  \ref{fig-3d2}(b), 
respectively, to show the regions hidden in fig.\ref{fig-3d1}(a) and 
 \ref{fig-3d2}(a) behind the surfaces I, II and III. 

By drawing flat surfaces parallel to the $\alpha$-$\beta$ plane, each 
corresponding to a fixed value of $P_{\omega_{a}}$ (in (a)) or 
$P_{\omega_{h2}}$ (in (b)), we have obtained 
the curves of intersection of this flat plane with the surfaces I, II 
and III. By projecting these curves on the plane  $P_{\omega_{h2}} = 0$, 
we also obtained the 2d phase diagram of the system in the $\alpha$-$\beta$ 
plane for several different values of $P_{\omega_{h2}}$. This phase 
diagram helps in comparing and contrasting our results for the ribosome 
traffic model with the 2d phase diagram of the TASEP in the  
$P_{\omega_{h2}}$ plane (Fig.\ref{fig-2d}). The most interesting feature 
is that, unlike TASEP, the lines on which HD and LD phases coexist are 
curved. This characteristic seems to be the general feature of such phases 
diagrams, rather than an exception; similar curved lines of coexistence 
between HD and LD phases have been observed also in some other contexts 
\cite{antal00}.

\subsubsection{Average density profiles}

The bulk density of the system is governed by the following equations:
\begin{widetext}
\begin{eqnarray}
\rho = \left\{ \begin{array}{lll}
         \rho_{-}&~ \mbox{if}~~ \beta~ > ~\frac{\alpha~\ell~P_{\omega_{h2}}}{(1+\Omega_{h2})(P_{\omega_{h2}}-\alpha+\ell \alpha-\Omega_{h2}\alpha)}~~~\mbox{and}~ \alpha < \frac{P_{\omega_{h2}}\rho_{*}}{(1+\Omega_{h2})(1-\rho_{*}(\ell-1))}\Rightarrow \mbox{LD}\\
           \rho_{+}&~~ \mbox{if}~~ \beta < \frac{P_{\omega_{h2}}(1-\rho_{*}~\ell)}{(1+\Omega_{h2})(1-\rho_{*}(\ell -1))} ~~~\mbox{and}~~ \alpha > \frac{P_{\omega_{h2}}\beta(1+\Omega_{h2})}{P_{\omega_{h2}}\ell+\beta(1-\ell+2\Omega_{h2}-\ell
\Omega_{h2}+\Omega^2_{h2})}\Rightarrow \mbox{HD}\\
\rho_{*}&~\mbox{if}~~\beta > \frac{P_{\omega_{h2}}(1-\rho_{*}~\ell)}{(1+\Omega_{h2})(1-\rho_{*}(\ell -1))}~~\mbox{and}~~\alpha > \frac{P_{\omega_{h2}}\rho_{*}}{(1+\Omega_{h2})(1-\rho_{*}(\ell-1))}\Rightarrow\mbox{MC}.\end{array} \right.
\end{eqnarray}
\end{widetext}

\section{Summary and conclusion}\label{summary}

In this paper we have derived the exact analytical expression for 
the distribution of the dwell times of ribosomes at each codon on 
the mRNA track. For this purpose we have used a model that captures 
the essential steps in the mechano-chemical cycle of a ribosome. 
As more details of this cycle get unveiled by new experiments, our 
model can be extended to capture those new features and the dwell 
time distribution can be re-calculated accordingly. Moreover, some of 
the transitions in the mechano-chemical cycle used in our model may 
require reinterpretation to reconcile with new observations. 
Neverthless, at this stage, the dwell time distribution predicted 
by our theory agrees qualitatively with the corresponding distribution 
observed {\it in-vitro} single ribosome experiments. Moreover, our 
prediction can be tested quantitatively by repeating the single 
ribosome experiments varying the supply of amino acid monomers and GTP 
molecules. 

From the full distribution, we have also calculated the mean dwell 
time which satisfies a Michaelis-Menten-like equation. We have 
pointed out the formal similarities between the cycles, and the 
corresponding equations, for a single enzyme molecule and a single 
ribosome, which are responsible for the Michalis-Menten-like form 
of the mean-dwell time. The inverse of the mean-dwell time is also 
the average velocity of the ribosome. The expression of this average 
velocity obtained from the dwell time distribution is identical to 
that obtained by an alternative approach pioneered by Fisher and 
Kolomeisky in the context of generic models of molecular motors. 
Finally, following standard procedure, we capture the effects of 
load force by modifying the rate constant $\omega_{h2}$ and predict 
the force-velocity relation and its dependence on experimentally 
controllable parameters. From this relation we have estimated 
the stall force of a ribosome. Our theoretical estimate is consistent 
with the experimentally measured value reported in the literature. 
However, to our knowledge, the full force-velocity relation for 
ribosomes has not been measured so far. But, with the rapid progress 
in the experimental techniques, it should be possible in near future 
to test the full force-velocity relation predicted by our theory.

We have presented a few quantitative characteristics of the 
fluctuations in the kinetics of ribosomes. We have defined a 
``randomness parameter'' $r$, which is a measure of the fluctuations 
in the dwell times. From the full probability density of the 
dwell times, we have derived the expression for $r$ and analyzed 
some of its interesting features. We have also reported the 
analytical expression for the diffusion constant and related it 
to the mean velocity and the randomness parameter. Using the 
central limit theorem, we have argued that the distribution of 
the run times of the ribosomes from the start codon to the stop 
codon is Gaussian and also pointed out the relations between 
its first two moments and those of the dwell time distribution.

To our knowledge, the run time distribution of ribosomes has not 
been measured so far. RNA polymerase (RNAP) motor runs on a DNA 
track using the track to polymerize the complementary RNA. There 
are some similarities between template-dictated polymerizations  
driven by ribosome and RNAP. The run time distribution of RNAP 
has been measured and found to be Gaussian \cite{tolic}. This is 
consistent with the Gaussian run time distribution for ribosomes 
predicted in this paper which follows from very general arguments 
based on the central limit theorem. 

Incorporating inter-ribosome steric interactions in the model, we 
have developed a model for ribosome traffic. The model may be 
regarded as a TASEP for hard rods each of which has five distinct 
``internal states''; transitions between these internal states 
constitute parts of the mechano-chemical cycle of a ribosome. 
Initiation and termination of the polymerization of individual 
proteins are captured by imposing open boundary conditions. 
For this model, we have drawn three-dimensional phase diagrams in 
spaces spanned by parameters which can be varied in a controlled 
manner in laboratory experiments {\it in-vitro}. In principle, 
the phase diagram can be obtained by analyzing the density profile 
of the ribosomes in electron micrographs of the system for several 
different concentrations of amino acid subunits, GTP concentration etc. 

\noindent{\bf Appendix I} 

\begin{figure}
 \begin{center}
\includegraphics[width=0.9\columnwidth]{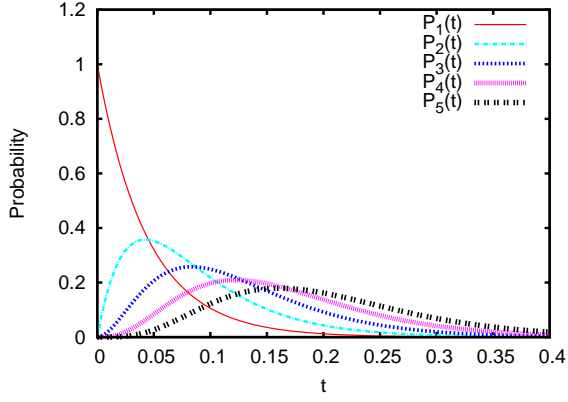}
\end{center}
\caption{(Color online) The probabilities $P_{\mu}(t)$ 
($\mu=1,2,...,5$), corresponding to the initial conditions 
$P_1(0)=1$ and $P_2(0)=P_3(0)=P_4(0)=P_5(0)=0$ (see equations 
(\ref{eq-fullp1})-(\ref{eq-fullp5})) are plotted for the parameter 
set $\omega_a = 23.0$, $\omega_{h1} = 24.0$, $k_2 = 25.0$, 
$\omega_g = 26.0$, $\omega_{h2} = 27.0$ and $\omega_{p} = 0.5$.}
\label{fig-p1top5}
\end{figure}

The solution of the equations (\ref{eq-m1})(\ref{eq-m6}), 
for the initial condition (\ref{6}) are given by 
\begin{widetext}
\begin{equation}
P_1(t)= exp(-\omega_a t) + 
\omega_{a} \omega_{p}\biggl[\frac{exp(-\omega_1 t)}{(\omega_a-\omega_1)(\omega_2-\omega_1)} + \frac{exp(-\omega_2 t)}{(\omega_a-\omega_2)(\omega_1-\omega_2)} + \frac{exp(-\omega_a t)}{(\omega_1-\omega_a)(\omega_2-\omega_a)}\biggr] 
\label{eq-fullp1}
\end{equation}
\begin{equation}
P_2(t)=\omega_{a}\biggl[\frac{exp(-\omega_1 t)}{(\omega_2-\omega_1)}+\frac{exp(-\omega_2 t)}{(\omega_1-\omega_2)}\biggr]
\label{eq-fullp2}
\end{equation}
\begin{equation}
P_3(t)=\omega_{a}\omega_{h1}\biggl[\frac{exp(-\omega_1 t)}{(\omega_2-\omega_1)(k_2-\omega_1)}+\frac{exp(-\omega_2 t)}{(\omega_1-\omega_2)(k_2-\omega_2)}+\frac{exp(-k_2 t)}{(\omega_1-k_2)(\omega_2-k_2)}\biggr]
\label{eq-fullp3}
\end{equation}
\begin{eqnarray}
P_4(t)=\omega_{a}\omega_{h1}k_2&\biggl[&\frac{exp(-\omega_1 t)}{(\omega_2-\omega_1)(k_2-\omega_1)(\omega_g-\omega_1)}+\frac{exp(-\omega_2 t)}{(\omega_1-\omega_2)(k_2-\omega_2)(\omega_g-\omega_2)} \nonumber \\ 
&+&\frac{exp(-k_2 t)}{(\omega_1-k_2)(\omega_2-k_2)(\omega_g-k_2)}+\frac{exp(-\omega_g t)}{(\omega_1-\omega_g)(\omega_2-\omega_g)(k_2-\omega_g)}\biggr]
\label{eq-fullp4}
\end{eqnarray}
\begin{eqnarray}
P_5(t)=\omega_{a}\omega_{h1}k_2\omega_g&\biggl[&\frac{exp(-\omega_1 t)}{(\omega_2-\omega_1)(k_2-\omega_1)(\omega_g-\omega_1)(\omega_{h2}-\omega_1)}+\frac{exp(-\omega_2 t)}{(\omega_1-\omega_2)(k_2-\omega_2)(\omega_g-\omega_2)(\omega_{h2}-\omega_2)} \nonumber \\ 
&+&\frac{exp(-k_2 t)}{(\omega_1-k_2)(\omega_2-k_2)(\omega_g-k_2)(\omega_{h2}-k_2)}+\frac{exp(-\omega_g t)}{(\omega_1-\omega_g)(\omega_2-\omega_g)(k_2-\omega_g)(\omega_{h2}-\omega_g)} \nonumber \\
&+&\frac{exp(-\omega_{h2} t)}{(\omega_1-\omega_{h2})(\omega_2-\omega_{h2})(k_2-\omega_{h2})(\omega_g-\omega_{h2})}\biggr]
\label{eq-fullp5}
\end{eqnarray}
\end{widetext}
These distributions are plotted in fig.\ref{fig-p1top5} for one set 
of values of the model parameters. These clearly shows that the 
probability $P_{1}(t)$ decreases monotonically from the initial 
value $1$ while the states $2$, $3$, $4$ and $5$ ``rise'' and 
``fall'' in a sequence.

\noindent {\bf Acknowledgements}: 
We thank G. M. Sch\"utz for useful correspondences. This work has been 
supported by a grant from CSIR (India). Debanjan Chowdhury and T.V. 
Ramakrishnan (TVR) thank DST, government of India, for a KVPY fellowship 
and a Ramanna fellowship, respectively. AG thanks UGC (India) for a 
senior research fellowship. TVR would also like to thank 
National Centre for Biological Sciences, Bangalore, for hospitality.

\vspace{1cm}


\end{document}